\documentclass[twoside]{article}

\usepackage{float}
\usepackage{color}
\usepackage{greekbf}
\usepackage{float}
\usepackage{amsmath,amssymb}
\usepackage{mathrsfs}
\usepackage{enumerate}
\usepackage{srcltx}
\usepackage{mathtools}
    \mathtoolsset{showonlyrefs}
\usepackage{graphicx, psfrag}
\usepackage{setspace}
\usepackage{hyperref}
\hypersetup{bookmarks=false,
colorlinks=false,
linkcolor=black,
urlcolor=black,
pagebackref=true,
pdfstartview={FitH},
bookmarksopenlevel=2
}

\newtheorem{remark}{Remark}

\DeclareMathAlphabet{\mathbf}{OT1}{cmr}{bx}{it}
\definecolor{red}{rgb}{0.9,0,0}
\definecolor{blue}{rgb}{0.2,0.2,0.8}
\definecolor{green}{rgb}{0.0,0.5,0.2}
\definecolor{darkblue}{rgb}{0.2,0.2,0.5}
\definecolor{orange}{rgb}{1,0.5,0}
\definecolor{pink}{rgb}{0.96,0.5,0.46}
\definecolor{lblue}{rgb}{0.18,0.74,1}
\definecolor{cyan}{rgb}{0,0.8,0.8}

\newcommand {\eb}{\mathbf{e}}
\newcommand {\gb}{\mathbf{g}}
\newcommand {\pb} {\mathbf{p}}
\newcommand {\qb} {\mathbf{q}}
\newcommand {\rb} {\mathbf{r}}
\newcommand {\ub} {\mathbf{u}}
\newcommand {\xb} {\mathbf{x}}

\newcommand {\Hb} {\mathbf{H}}

\newcommand {\nb} {\mathbf{n}}

\newcommand {\ab} {\mathbf{a}}

\newcommand {\Cc}  {\mathcal{C}}
\newcommand {\Dc}  {\mathcal{D}}

\newcommand {\Uc}  {\mathcal{U}}

\newcommand {\Zc}  {\mathcal{Z}}
\newcommand {\cb} {\mathbf{c}}
\newcommand {\db} {\mathbf{d}}
\newcommand {\bb} {\mathbf{b}}

\DeclareMathOperator{\Thz}{\overset{(z)}{\Theta}}
\DeclareMathOperator{\Thc}{\overset{(c)}{\Theta}}
\DeclareMathOperator{\ufb}{\boldsymbol{\mathfrak{u}}}

\begin{document}
	
\doublespacing

\title{\vspace{-3cm} {\bf  The Gaussian Stiffness of  Graphene \\  deduced from a Continuum Model \\ based on Molecular Dynamics Potentials}}

\author{
Cesare Davini$^1$\!\!\!\!\! \and  Antonino Favata$^2$\!\!\!\!\! \and Roberto Paroni$^{3}$
}

%
%\date{\today}

%%% ----------------------------------------------------------------------
\maketitle
%%% ----------------------------------------------------------------------

\vspace{-1cm}
\begin{center}
	{\small
		$^1$ Via Parenzo 17, 33100 Udine\\
		\href{mailto:cesare.davini@uniud.it}{cesare.davini@uniud.it}\\[8pt]
		$^2$ Department of Structural and Geotechnical Engineering\\
		Sapienza University of Rome, Rome, Italy\\
		\href{mailto:antonino.favata@uniroma1.it}{antonino.favata@uniroma1.it}\\[8pt]

		$^3$ Dipartimento di Architettura, Design e Urbanistica\\
		University of Sassari, Alghero (SS), Italy\\
		\href{mailto:paroni@uniss.it}{paroni@uniss.it}
	}
\end{center}

\pagestyle{myheadings}
\markboth{C.~Davini, A.~Favata, R.~Paroni }
{The Bending Behavior of  Graphene}

\vspace{-0.5cm}
\section*{Abstract}

We consider a discrete model of a graphene sheet with atomic interactions governed by a harmonic approximation of the 2nd-generation Brenner potential that depends on bond lengths, bond angles, and two types of dihedral angles. A continuum limit is then deduced that fully describes the bending behavior. In particular, we deduce for the first time an analytical expression of the \textit{Gaussian stiffness}, a scarcely investigated parameter  ruling the rippling of graphene,  for which contradictory values have been proposed in the literature. We disclose the atomic-scale sources of both bending and Gaussian stiffnesses and provide for them quantitative evaluations.

\vspace{1cm}
\noindent {\bf Keywords}: Graphene, Continuum Modeling, Gaussian stiffness.

\tableofcontents

\section{Introduction}\label{sec:INTROD}

Graphene   has attracted increasing interest during the past few years,  and is nowadays used in a great variety of applications, taking advantage of its extraordinary mechanical, electrical and thermal conductivity properties. Nevertheless, its potentialities, and those of graphene-based materials, are far from being fully explored and exploited, and many studies are carried out by the scientific community in order to develop new technological applications \cite{Ferrari2014}.

The understanding of the bending behavior of graphene is  of paramount importance in several technological applications. It is exploited, for example, to predict the performance of graphene nano-electro-mechanical devices and ripple formation \cite{Huang2006,Lu_2009,Zhang_2011,Lindahl_2012,Tapaszto_2012,Kim_2012,Shi_2012,Hartmann_2013,Hajgato_2012,Wei_2013,Pacheco_2014,Favata_2016b}, and it is proposed to be the key point to produce efficient hydrogen-storage devices \cite{Tozzini_2011,Goler_2013,Tozzini_2013}. 
A very recent review in \textit{Materials Today}  by  Deng \&  Berry \cite{Deng_2016} gives an overview on the hot problem of wrinkling, rippling and crumpling of graphene, highlighting formation mechanism and applications. Indeed, these corrugations  can modify its electronic structure, create polarized carrier puddles, induce pseudo-magnetic field in bilayers and alter surface properties. Although a great effort has been done on the experimental side, predictive models are still wanted. They are of crucial importance when these phenomena need to be controlled and designed.

In particular, since the bending stiffness and the Gaussian  stiffness ---that is the reluctance to form non-null Gaussian curvatures--- are the two crucial parameters  governing the rippling of  graphene, it is necessary to accurately determine them for both the design and the manipulation of graphene morphology. Although  several evaluations of the bending stiffness have been proposed in the literature, the Gaussian stiffness has not been object of an intensive study. Indeed, as pointed out in a very recent review on mechanical properties of graphene \cite{Akiwande_2016}, only two conflicting evaluations have been proposed. In \cite{Koskinen_2010}, periodic boundary conditions have been used within a quantum-mechanical framework, and the value $-0.7$ eV has been found. While  in \cite{Wei_2013} the estimate of $-1.52$ eV has been obtained by combining the configurational energy of membranes determined by Helfrich Hamiltonian with energies of fullerenes and single wall carbon nanotubes calculated by Density Functional Theory (DFT). At a discrete level there are  two main  difficulties in the evaluation of the Gaussian stiffness: on the one hand, controlling a discrete double curvature surfaces is problematic, and on the other hand, a suitable notion of Gaussian curvature at the discrete level should be introduced. Instead, when well established continuum models are adopted, such as plate theory,  one has the problem of determining the equivalent stiffnesses, letting alone the conceptual crux of giving a meaning to  the notion of thickness (see \cite{Huang2006}, \cite{Bajaj_2013} and references therein).

In this paper we deduce  a continuum 2-dimensional model of a graphene sheet inferred from Molecular Dynamics (MD). In particular, looking at the 2nd-generation reactive empirical bond-order (REBO) potential \cite{Brenner_2002}, we give a nano-scale description of the atomic interactions and then we deduce the continuum limit, avoiding the problem of postulating an ``equivalent thickness'' and circumventing artificial procedures to  identify the material parameters that describe the mechanical response of a  plate within the classical theory.

Our analysis of the atomic-scale interaction relies on the discrete mechanical model proposed in \cite{Favata_2016} and exploited in  \cite{Favata_2016a,Favata_2016b,Alessi_2016}, whose results are also based on the  2nd-generation Brenner potential. This potential is largely used in MD simulations for carbon allotropes; for a detailed description of its general form  and that adopted in our theory we  refer the reader to  Appendix B of \cite{Favata_2016}. Here, we recall the key ingredients needed:
\begin{enumerate}[(i)]
	\item the kinematic variables associated with the interatomic bonds involve first, second and third nearest neighbors of any given atom. In particular, the kinematical variables we consider are \textit{bond lengths}, \textit{bond angles}, and \textit{dihedral angles}; from \cite{Brenner_2002} it results that these latter are of two kinds, that we here term C and Z, as carefully described in Sec. \ref{sec:KIN}.
	\item graphene suffers an angular \textit{self-stress},  and the \textit{self-energy} associated with the self-stress (sometimes called \textit{cohesive energy} in the literature) is quantitatively relevant;
	\item the energetic contribution of dihedral interaction is very relevant in bending.
\end{enumerate}

For the first time, we propose a continuum model able to predict both the \textit{bending}  and the \textit{Gaussian stiffnesses}. The analytical formula we obtain for the former predicts exactly the same value as that computed with MD simulations of the last generation. The value of the Gaussian stiffness we obtain is in very good agreement with DFT computations proposed in \cite{Wei_2013}. 

For the modeling of graphene many different approaches at different scales can be found in the literature, ranging from first principle calculations \cite{Kudin_2001,Liu_2007}, atomistic calculations \cite{Zakharchenko_2009,Zhao_2009,SakhaeePour_2009} and continuum mechanics \cite{Cadelano_2009,Yakobson_1996,Luhuang2009,Scarpa_2009,Scarpa_2010,Sfyris_2014b,Sfyris_2014}. Furthermore, mixed atomistic formulations with finite elements have been reported for graphene \cite{Arroyo_2002,Arroyo2004}.

The paper is organized as follows. In Sec. \ref{sec:KIN}, we describe the kinematics and the energetic of the  graphene sheet at the nano-scale. In Sec. \ref{sec:STRAINS}, we deduce the  strain measures  for the change of edge lengths, wedge angles and dihedral angles, approximated to the lowest order that makes the energy quadratic in the displacement. In Sec. \ref{SEC:splitting}, the total energy is split in its in-plane and out-of-plane contributions, and focus is set on the latter, having the first already been considered in \cite{Davini_2014}. In Sec. \ref{SEC:continuum}, we deduce a continuous energy that approximates the discrete energy and in Sec. \ref{plate} the limit energy is rearranged in a more amenable form, able to put in evidence the equivalence with plate theory.  In Sec. \ref{num}, quantitative results for the continuum material parameters are deduced, by means of the 2nd-generation Brenner potential, and  compared with the literature.  Appendix \ref{appendix}, containing some computations ancillary to Sec. \ref{sec:STRAINS}, completes the paper.

\section{Description of kinematics and energetics of the graphene sheet} \label{sec:KIN}

\begin {figure} [!hbtp]
\begin{center}
	\includegraphics [height=6.5 cm] {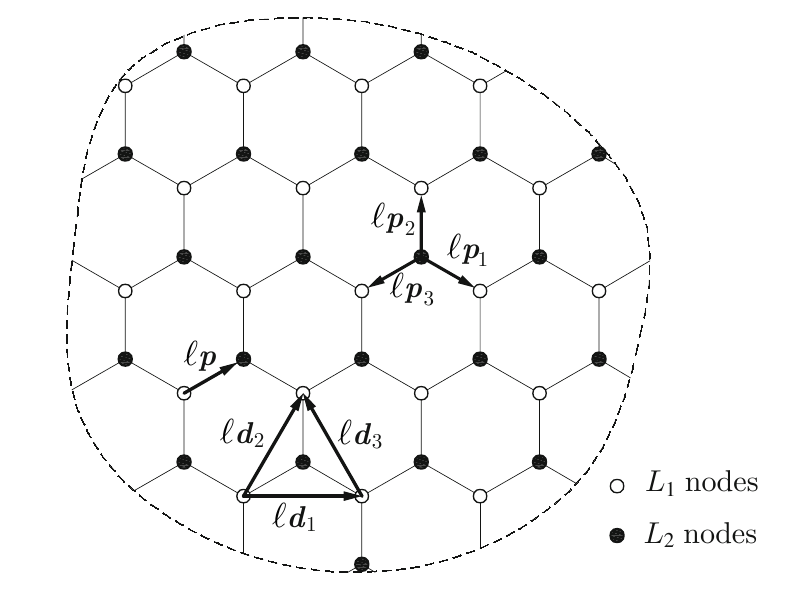}
\end{center}
\caption{The hexagonal lattice}\label{fg:LATTICE}
\end {figure}

At the nano-scale a graphene sheet is a discrete set of carbon atoms that, in the
absence of external forces, sit at the vertices of a
periodic array of hexagonal cells. More specifically,
atoms occupy the nodes of the $2$--{\it lattice}, see Figure~\ref{fg:LATTICE},  generated by two simple \textit{Bravais} lattices
\begin{equation}\label{eq:KIN_1}
\begin{array}{l} L_1(\ell) = \{ \mathbf{x} \in \mathbb{R}^2:
\mathbf{x} = n^1\ell \db_{1} + n^2\ell \db_{2} \quad \mbox{with} \quad (n^1,
n^2) \in \mathbb{Z}^2 \}, \\ L_2(\ell) = \ell\mathbf{p} + L_1(\ell),
\end{array}
\end{equation}
simply shifted with respect to one another. In \eqref{eq:KIN_1}, $\ell$ denotes
the lattice size (the reference {\it interatomic distance}), while
$\ell\db_{\alpha}$ and $\ell\mathbf{p}$ respectively are the {\em lattice
	vectors} and the {\em shift vector}, whose Cartesian components
are given by
\begin{equation}\label{eq:KIN_2}
\db_{1} =(\sqrt{3}, 0), \quad \db_{2} = (\frac{\sqrt{3}}{2},
\frac{3}{2} ) \quad \mbox{and} \quad \mathbf{p} =
(\frac{\sqrt{3}}{2}, \frac{1}{2}).
\end{equation}
The sides of the hexagonal cells in Figure~\ref{fg:LATTICE} stand for the bonds
between pairs of next nearest neighbor atoms and are represented by the
vectors
\begin{equation}\label{eq: KINENER 3}
\pb_\alpha = \db_\alpha - \pb \ \  (\alpha = 1, 2) \quad \mbox{and}
\quad \pb_3 = - \, \pb.
\end{equation}
For convenience we also set $$\db_3=\db_2-\db_1.$$

As reference configuration we take the set of points $\xb^\ell\in L^1(\ell)\cup L^2(\ell)$ contained in a bounded open set $\Omega\subset \mathbb{R}^2$.

\begin{figure}[h]
	\centering
	\includegraphics[scale=1]{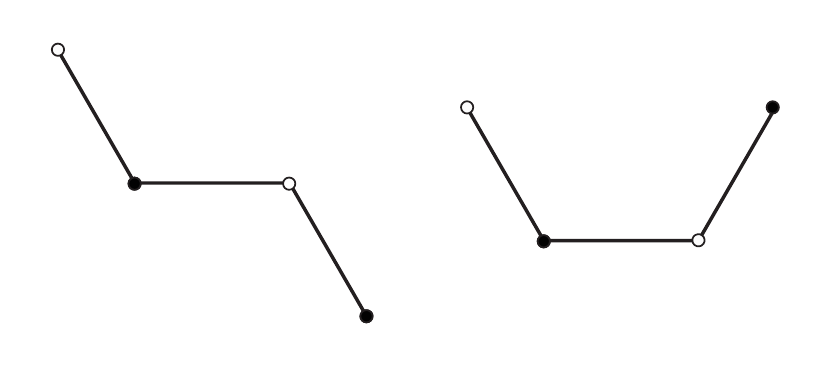}
	\caption{A Z-dihedral angle (left) and a C-dihedral angle (right).}
	\label{fig:bondchain}
\end{figure}

Graphene mechanics is ruled by the interactions between
the carbon atoms given by some suitable potential. According to the 2nd-generation Brenner potential \cite{Brenner_2002}, as detailed in \cite{Favata_2016, Favata_2016a}, in order to account properly for the mechanical behavior of a bended graphene sheet it is necessary to consider three types of energetic contributions, respectively coming from: binary interactions between next nearest atoms  ({\it edge bonds}), three-bodies interactions between consecutive pairs of next nearest atoms  ({\it wedge bonds}) and four-bodies interactions between three consecutive pairs of next nearest atoms ({\it dihedral bonds}).  There are two types of relevant dihedral bonds: the Z-dihedra
in which the edges connecting the four atoms form a z-shape and the C-dihedra in which the edges form a c-shape (see Fig.\ \ref{fig:bondchain}).

\begin{figure}[h]
	\centering
	\includegraphics[scale=1]{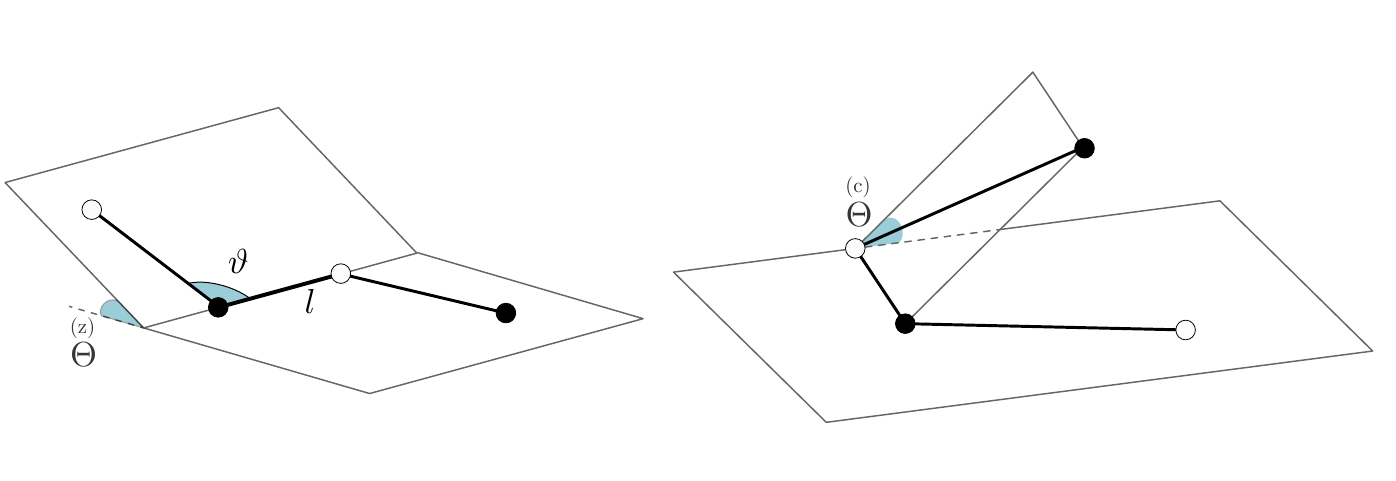}
	\caption{Kinematic variables: distance $l$, angle $\vartheta$, Z-dihedral angle $\Thz$ and C-dihedral angle $\Thc$.}
	\label{fig:bonds}
\end{figure}

We consider a harmonic approximation of the stored energy and assume that it is given by the sum of the following terms:

\begin{equation}\label{eq:ENER 1}
\begin{array}{l}
\displaystyle \mathcal{U}_\ell^l= \frac{1}{2} \, \sum_{\mathcal {E}} k^l \, (l  -
l^{\mbox{\scriptsize nat}})^2 ,\\
\displaystyle\mathcal{U}_\ell^\vartheta = \frac{1}{2} \, \sum_{\mathcal {W}} k^\vartheta \, (\vartheta
\, - \, \vartheta^{\mbox{\scriptsize nat}})^2, \\
\displaystyle\mathcal{U}_\ell^\Theta = \frac{1}{2} \, \sum_{\mathcal {Z}} k^\Zc \, (\Thz
\, - \, \Theta^{\mbox{\scriptsize nat}})^2+\frac{1}{2} \, \sum_{\mathcal {C}} k^\Cc \, (\Thc
\, - \, \Theta^{\mbox{\scriptsize nat}})^2
\end{array}
\end{equation}
$\mathcal{U}_\ell^l$, $\mathcal{U}_\ell^\vartheta $ and $\mathcal{U}_\ell^\Theta$ are the energies of the edge bonds, the wedge bonds and the dihedral bonds, respectively; $l$ denotes the distance between nearest neighbor atoms, $\vartheta$ the angle between pairs of edges having a lattice point in common and $\Thz$ and $\Thc$ the Z- and C-dihedral angles between two consecutive wedges, to be  defined later (see Fig. \ref{fig:bonds}); $l^{\mbox{\scriptsize nat}} $ is the edge length at ease, 
$\vartheta^{\mbox{\scriptsize nat}}$ the angle  at ease between consecutive edges and 
$\Theta^{\mbox{\scriptsize nat}}$ the dihedral angle at ease. 

The sums extend to
all edges, $\mathcal{E}$, all wedges, $\mathcal{W}$, all Z-dihedra, $\mathcal{Z}$, and all C-dihedra, $\Cc$, contained in the set $\Omega$. The
bond constants $k^l$, $k^\vartheta$, $k^\Zc $,  and $k^\Cc$  will be deduced by making use of the 2nd-generation Brenner potential.

The graphene sheet does not have a configuration at ease ({\it i.e.} stress-free). Indeed, in \cite{Favata_2016} it has been shown that 
$$
\Theta^{\mbox{\scriptsize nat}}=0, \quad l^{\mbox{\scriptsize nat}} =\ell \quad \mbox{and} \quad 
\vartheta^{\mbox{\scriptsize nat}}=\frac{2}{3}\pi + \delta\vartheta_0,$$
where $\delta\vartheta_0\ne 0$.
We set $\delta\Theta:=\Theta$, $l= \ell + \delta l$ and $\vartheta=\frac{2}{3}\pi + \delta\vartheta$ and write \eqref{eq:ENER 1} as
\begin{equation}\label{eq:ENER 1bis}
\begin{array}{l}
\displaystyle \mathcal{U}_\ell^l = \frac{1}{2} \, \sum_{\mathcal {E}} k^l \, (\delta l)^2 ,\\
\displaystyle\mathcal{U}_\ell^\vartheta = \frac{1}{2} \, \sum_{\mathcal {W}} k^\vartheta \, (\delta\vartheta
\, - \, \delta\vartheta_0)^2, \\
\displaystyle\mathcal{U}_\ell^\Theta = \frac{1}{2} \, \sum_{\mathcal {Z}} k^{\Zc}\, (\delta\Thz)^2+\frac{1}{2} \, \sum_{\mathcal {C}} k^{\Cc}\, (\delta\Thc)^2
\end{array}
\end{equation}\label{eq:ENER 1bis_1}
In particular, up to a constant, the wedge energy takes the form
\begin{equation}\label{utth}
\mathcal{U}_\ell^\vartheta = \tau_{0} \sum_{\mathcal {W}}\delta \vartheta +  \frac{1}{2} \, \sum_{\mathcal {W}} k^\vartheta \, (\delta\vartheta)^2,
\end{equation}
with 
\begin{equation}\label{deft0}
\tau_0 := -k^\vartheta \, \delta\vartheta_0
\end{equation} 
the {\it angle self-stress}.
The dihedral bonds play an important role because they contribute to the stored energy by about 50\%, see \cite{Favata_2016,Favata_2016b}, the rest is due to the angle self-stress $\tau_0$ associated to the wedge bonds. 

The energy decomposition \eqref{eq:ENER 1bis} is based on the choice of the set of kinematical variables $\{l, \vartheta, \Thz, \Thc\}$. This choice is the most natural, if one considers the 2nd-generation Brenner potential, where all those variables  appear in explicit manner.  A harmonic approximation  in each of those parameters is of course unique. 	

In the next section we shall make explicit the change of length $\delta l$, the change of wedge angle $\delta \vartheta$, and the changes of the Z- and C-dihedral angles $\delta\Thz$ and $\delta\Thc$.
In Section \ref{SEC:splitting}, with the notation introduced in the next section, we shall write the energies
\eqref{eq:ENER 1bis} more explicitly.

\section{Approximated strain measures}\label{sec:STRAINS}
In this section we calculate the strain measures associated to a change of configuration described by a displacement field $\ub : (L_1(\ell)\cup L_2(\ell))\cap \Omega \to\mathbb{R}^3$, approximated to the lowest order that makes the energy quadratic in $\ub$.

\subsection{Change of the edge lengths}
With $\delta l_i(\xb^\ell)$ we denote the change in length of the edge parallel to $\pb_i$ and starting from the lattice point $\xb^\ell\in (L_1(\ell)\cup L_2(\ell))\cap \Omega$. Thus,
\begin{align*}
\delta l_i(\xb^\ell)&=|(\xb^\ell + \ell\pb_i + \ub(\xb^\ell+\ell\pb_i)) - (\xb^\ell +\ub(\xb^\ell))|-\ell\\
&=|\ell\pb_i + (\ub(\xb^\ell+\ell\pb_i) - \ub(\xb^\ell))|-\ell,
\end{align*}
and up to terms $o(|\ub |)$ can be rewritten as
\begin{equation}\label{eq:delta_l}
\delta l_i (\xb^\ell)= \frac{1}{\ell}\,(\ub(\xb^\ell+\ell\pb_i) - \ub(\xb^\ell))\cdot\mathbf{p}_i \qquad i=1,2,3.
\end{equation}
In particular, the first order changes are determined by the in-plane components of $\ub$ only.

\subsection{Change of the wedge angles}\label{sec:WANGLE}
\begin{figure}[H]
	\centering
	\includegraphics[scale=1]{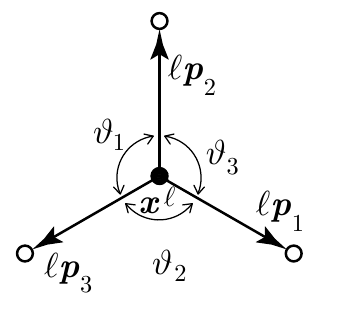}
	\caption{The wedge angles $\vartheta_i$.}
	\label{fig:angles_app}
\end{figure}

For each fixed node $\xb^\ell\in (L_1(\ell)\cup L_2(\ell))\cap \Omega$ we denote by $\vartheta_i(\xb^\ell)$ the angle of the wedge delimited by the edges $\pb_{i+1}$ and $\pb_{i+2}$; that is, the wedge angle opposite to the $i$-th edge (see Fig. \ref{fig:angles_app}). Here, $i, i+1$, and $i+2$ take values in $\{1,2,3\}$ and the sums should be interpreted mod 3: for instance, if $i=2$ then $i+1=3$ and $i+2=1$.
From \eqref{utth} we see that the change in the wedge angle enters into the energy not just quadratically but also linearly, therefore the variations of the wedge angle should be computed up to the second order approximation.
To keep the notation compact, we set
$$
\ub_{i}:=\ub(\xb^\ell+\ell\pb_{i}),\quad\mbox{and}\quad \ub_0:=\ub(\xb^\ell).
$$
Let 
\begin{align*}
\qb_{i+1}&:=(\xb^\ell + \ell\pb_{i+1} + \ub(\xb^\ell+\ell\pb_{i+1})) - (\xb^\ell +\ub(\xb^\ell))\\&=\ell\pb_{i+1}+(\ub_{i+1} -\ub_0),
\end{align*}
and
\begin{align*}
\qb_{i+2}&:=(\xb^\ell + \ell\pb_{i+2} + \ub(\xb^\ell+\ell\pb_{i+2})) - (\xb^\ell +\ub(\xb^\ell))\\&=\ell\pb_{i+2}+(\ub_{i+2} -\ub_0),
\end{align*}
be the images of the edges parallel to $\pb_{i+1}$ and $\pb_{i+2}$ and starting at $\xb^\ell$. Then, the angle $\vartheta_i=\vartheta_i(\xb^\ell)$ is given by
\begin{equation}\label{eq:wangle}
\cos(\vartheta_i)=\frac{{\qb_{i+1}}\cdot{\qb_{i+2}}}{|{\qb_{i+1}}||{\qb_{i+2}}|}.
\end{equation}

Calculations given in Appendix~\ref{app:WANGLE} yield that
\begin{equation}\label{eq:wangle_1}
\vartheta_i= \frac{2}{3}\pi+\delta\vartheta_i^{(1)}+\delta\vartheta_i^{(2)} + o(|\ub |^2),
\end{equation}
where $\delta\vartheta_i^{(1)}$ and $\delta\vartheta_i^{(2)}$ are the first order and the second order variation, respectively, of the wedge angle with respect to the reference angle $\frac{2}{3}\pi$. Therefore, keeping up to second order terms one has that
\begin{equation}\label{eq:wangle_1bis}
\delta\vartheta_i=\delta\vartheta_i^{(1)}+\delta\vartheta_i^{(2)}.
\end{equation}
It turns out that the first order variation takes the form\begin{equation}\label{eq:wangle1st_bis}
\delta\vartheta_i^{(1)}(\xb^\ell)=-\frac{1}{\ell}(\ub_{i+1}-\ub_0)\cdot\pb_{i+1}^\perp  + \frac{1}{\ell}(\ub_{i+2} -\ub_0)\cdot\pb_{i+2}^\perp,
\end{equation}
with  $\pb_{i+1}^\perp$ defined by
\begin{equation}\label{piperp}
\pb_{i+1}^\perp:=\frac{\pb_{i+2} +\frac{1}{2}\pb_{i+1}}{|\pb_{i+2} +\frac{1}{2}\pb_{i+1}|}=\frac{2}{\sqrt{3}}(\pb_{i+2} +\frac{1}{2}\pb_{i+1}) \qquad i=1, 2, 3,
\end{equation}
that is, the unit vector orthogonal to $\pb_{i+1}$, (cf. equation (19) in \cite{Favata_2014}).
\begin{figure}[h!]
	\centering
	\includegraphics[scale=1]{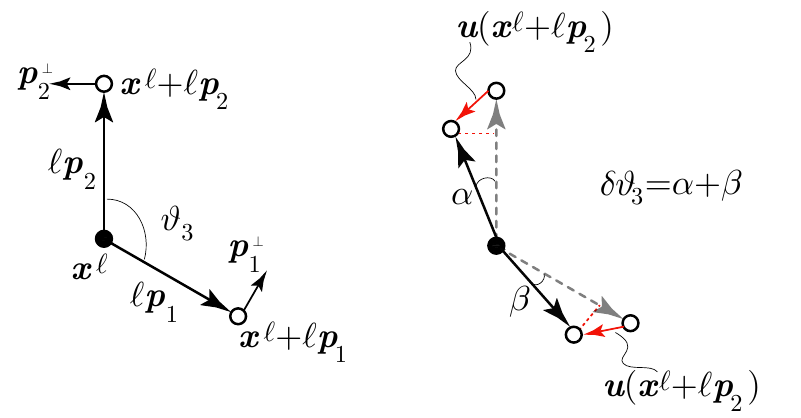}% "%" necessario
	\caption{The first order approximation of the change in the wedge angle.}\label{wangle}
\end{figure}
Figure~\ref{wangle} illustrates the geometrical meaning of formula \eqref{eq:wangle1st_bis}. 
In particular, 
$$
\sum_{i=1}^3\delta\vartheta_i^{(1)}(\xb^\ell)=0,
$$
as it could have been deduced from geometrical considerations.

The second order variation is given by, see Appendix~\ref{app:WANGLE},
\begin{equation}\label{eq:deltatheta_bis}
\begin{aligned}
\delta\vartheta_i^{(2)}(\xb^\ell)=-\frac{1}{\sqrt{3}}\Bigg[&-\frac{1}{2}{(\delta\vartheta_i^{(1)})}^2+\frac{2}{\ell^2}(\mathbf{u}_{i+1} - \mathbf{u}_0)\cdot(\mathbf{u}_{i+2} - \mathbf{u}_0)-\\
&\left(\frac{(\mathbf{u}_{i+1} - \mathbf{u}_0)}{\ell^2}\cdot\pb_{i+1}+\frac{(\mathbf{u}_{i+2} - \mathbf{u}_0)}{\ell^2}\cdot\pb_{i+2}  \right)\times\\
&\times\left(\frac{(\mathbf{u}_{i+1} - \mathbf{u}_0)}{\ell^2}\cdot\pb_{i+2}+\frac{(\mathbf{u}_{i+2} - \mathbf{u}_0)}{\ell^2}\cdot\pb_{i+1} -\frac{\sqrt{3}}{2}\delta\vartheta_i^{(1)}  \right)
\\
& -\frac{\pb_{i+1}\cdot\pb_{i+2}}{\ell^4}\Big(|\mathbf{u}_{i+1} - \mathbf{u}_0|^2-2\frac{1}{\ell^2}\big( \pb_{i+1}\cdot(\mathbf{u}_{i+1} - \mathbf{u}_0) \big)^2
\\ 
&+|\mathbf{u}_{i+2} - \mathbf{u}_0|^2-2\frac{1}{\ell^2}\big( \pb_{i+2}\cdot(\mathbf{u}_{i+2} - \mathbf{u}_0) \big)^2    \Bigg].
\end{aligned}
\end{equation}
By algebric manipulation one finds that
\begin{equation}\label{eq:sumdth_1}
\sum_{i=1}^3\delta\vartheta_i(\xb^\ell)=\sum_{i=1}^3\delta\vartheta_i^{(2)}(\xb^\ell)=-\frac{3\sqrt{3}}{\ell^2}\Bigg(\frac 13 \sum_{i=1}^3w(\xb^\ell+\ell\pb_i)-w(\xb^\ell)\Bigg)^2,
\end{equation}
where $w$ denotes the out-of-plane component of the displacement, that is
$$
w:=\ub\cdot\eb_3,
$$
where $\eb_3$ is the unit vector perpendicular to the undeformed sheet.
Note that, by \eqref{eq:sumdth_1}, the $\sum_i\delta\vartheta_i(\xb^\ell)$ is non-positive and hence the contribution of the self-stress to the strain energy is non-negative for $\tau_0<0$, i.e., for $\delta \vartheta_0>0$, see \eqref{deft0}.

\subsection{Change of the dihedral angles}

\begin{figure}[h]
	\centering
	\includegraphics[scale=1]{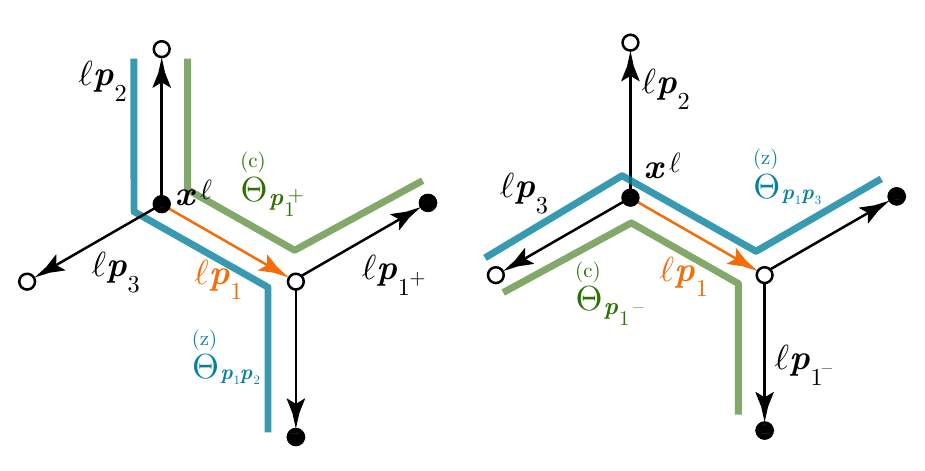}
	\caption{Left: C-dihedral angles $\Thc_{\pb_1^+}$ (green) and Z-dihedral angle $\Thz_{\pb_1\pb_2}$ (blue). Right: C-dihedral angles $\Thc_{\pb_1^-}$ (green) and Z-dihedral angle $\Thz_{\pb_1\pb_3}$ (blue).}
	\label{fig:cell_text}
\end{figure}

For each fixed node $\xb^\ell\in L_2(\ell)\cap \Omega$ and  for each edge parallel to $\pb_i$
and starting at $\xb^\ell$ we need to define four types of dihedral angles $\Thc_{\pb_i^+}(\xb^\ell), \Thc_{\pb_i^-}(\xb^\ell), \Thz_{\pb_i\pb_{i+1}}(\xb^\ell)$ and $\Thz_{\pb_i\pb_{i+2}}(\xb^\ell)$:

\begin{equation}
\begin{aligned}
&\cos\Thc_{\pb_i^+}=\frac{(\qb_i\times\qb_{i+1})\cdot(\qb_{i}\times\qb_{i^+})}{|\qb_i\times\qb_{i+1}||\qb_{i}\times\qb_{i^+}|},\\
&\cos\Thc_{\pb_i^-}=\frac{(\qb_{i+2}\times\qb_{i})\cdot(\qb_{i^-}\times\qb_{i})}{|\qb_{i+2}\times\qb_{i}||\qb_{i^-}\times\qb_{i})|},\\
&\cos\Thz_{\pb_i\pb_{i+1}}=\frac{(\qb_i\times\qb_{i+1})\cdot(\qb_{i^-}\times\qb_i)}{|\qb_i\times\qb_{i+1}||\qb_{i^-}\times\qb_i|},\\
& \cos\Thz_{\pb_i\pb_{i+2}}=\frac{(\qb_{i+2}\times\qb_i)\cdot(\qb_i\times\qb_{i^+})}{|\qb_{i+2}\times\qb_i||\qb_i\times\qb_{i^+}|},
\end{aligned}
\end{equation}
where	
\begin{equation}
\begin{aligned}
\qb_{i^+}=&\xb^\ell+\ell\pb_i-\ell\pb_{i+2}+\ub(\xb^\ell+\ell\pb_i-\ell\pb_{i+2})-\big( \xb^\ell+\ell\pb_i+\ub(\xb^\ell+\ell\pb_i) \big)\\
=&-\ell\pb_{i+2}+\ub_{i^+}-\ub_i, \qquad\qquad \ub_{i^+}:=\ub(\xb^\ell+\ell\pb_i-\ell\pb_{i+2}),\\
\qb_{i^-}=&\xb^\ell+\ell\pb_i-\ell\pb_{i+1}+\ub(\xb^\ell+\ell\pb_i-\ell\pb_{i+1})-\big( \xb^\ell+\ell\pb_i+\ub(\xb^\ell+\ell\pb_i) \big)\\
=&-\ell\pb_{i+1}+\ub_{i^-}-\ub_i, \qquad\qquad \ub_{i^-}:=\ub(\xb^\ell+\ell\pb_i-\ell\pb_{i+1})
\end{aligned}
\end{equation}
are the images of vectors $\ell\pb_{i^+}$ and $\ell\pb_{i^-}$ (see Fig. \ref{fig:cell_text}, for $i=1$), parallel to $\pb_{i+2}$ and $\pb_{i+1}$ and starting at the image of the point $\xb^\ell+\ell\pb_i$.

Also here, $i, i+1$, and $i+2$ take values in $\{1,2,3\}$ and the sums should be interpreted mod 3: for instance, if $i=3$ then $i+1=1$ and $i+2=2$.

The  C-dihedral angle $\Thc_{\pb_i^+}(\xb^\ell)$ is the angle corresponding to the C-dihedron with middle edge $\ell \pb_i$ and oriented as $\pb_i^\perp$,  while $\Thc_{\pb_i^-}(\xb^\ell)$ is the angle corresponding to the C-dihedron oriented opposite to $\pb_i^\perp$ (see Fig.~\ref{fig:cell_text} for $i=1$). 
The Z-dihedral angle $ \Thz_{\pb_i\pb_{i+1}}(\xb^\ell)$ 
corresponds to the Z-dihedron with middle edge $\ell \pb_i$ and the other two edges parallel to $\pb_{i+1}$
(see Fig.~\ref{fig:cell_text} for $i=1$).

\noindent 
Then, recalling that $\delta\Theta=\Theta$, calculations in Appendix~\ref{app:DANGLE} yield that
\begin{equation}\label{C-wedge}
\begin{aligned}
\delta\Thc_{\pb_i^+}(\xb^\ell)=\frac{2\sqrt{3}}{3\ell}[2w(\xb^\ell)-w(\xb^\ell+\ell \pb_{i+1})+w(\xb^\ell+\ell\pb_i-\ell \pb_{i+2})-2w(\xb^\ell+\ell \pb_{i})],
\end{aligned}
\end{equation}		
and
\begin{equation}\label{Z-wedge}
\begin{aligned}
\delta\Thz_{\pb_i\pb_{i+1}}(\xb^\ell)=
\frac{2\sqrt{3}}{3\ell}[w(\xb^\ell+\ell\pb_i-\ell \pb_{i+1})-w(\xb^\ell+\ell \pb_{i})+w(\xb^\ell+\ell \pb_{i+1})-w(\xb^\ell)].
\end{aligned}
\end{equation}		
Analogous formulas hold for $\delta\Thc_{\pb_i^-}$ and  $\delta\Thz_{\pb_i\pb_{i+2}}$:
\begin{equation}\label{CZ-wedge}
\begin{aligned}
&\delta\Thc_{\pb_i^-}(\xb^\ell)=-\frac{2\sqrt{3}}{3\ell}[2w(\xb^\ell)-w(\xb^\ell+\ell \pb_{i+2})+w(\xb^\ell+\ell\pb_i-\ell \pb_{i+1})-2w(\xb^\ell+\ell\pb_i)],\\
& \delta\Thz_{\pb_i\pb_{i+2}}(\xb^\ell)=\frac{2\sqrt{3}}{3\ell}[w(\xb^\ell+\ell\pb_i-\ell \pb_{i+2})-w(\xb^\ell+\ell \pb_{i})+w(\xb^\ell+\ell \pb_{i+2})-w(\xb^\ell)].
\end{aligned}
\end{equation}

\section{Splitting of the energy}\label{SEC:splitting}

The above calculations show that $\delta\vartheta_i^{(1)}$ as well as $\delta l_i$ depend upon the in-plane components of $\ub$, cf.\  \eqref{eq:delta_l} and \eqref{eq:wangle1st_bis}, while $\delta\vartheta_i^{(2)}$, $\delta\Thc$, and $\delta\Thz$ depend upon the out-of-plane component of $\ub$, cf.\  \eqref{eq:sumdth_1}, \eqref{C-wedge}, and \eqref{Z-wedge}. This yields a splitting of the energy into \textit{membrane} and \textit{bending} parts 
$$
\mathcal{U}_\ell=\mathcal{U}_\ell^{(m)}+\mathcal{U}_\ell^{(b)}, \quad \mathcal{U}_\ell^{(b)}:=\Uc_\ell^{(s)} +\Uc_\ell^{(d)} 
$$   
defined by
\begin{equation}
\begin{aligned}
& \mathcal{U}_\ell^{(m)}:=\frac{1}{2} \, \sum_{\mathcal {E}} k^l \, (\delta l)^2+\frac{1}{2} \, \sum_{\mathcal {W}} k^\vartheta \, {(\delta\vartheta^{(1)})}^2 \\
&\Uc_\ell^{(s)} :=\tau_{0} \sum_{\mathcal {W}}\delta \vartheta^{(2)}, \\ 
& \Uc_\ell^{(d)} :=\frac{1}{2} \, \sum_{\mathcal {Z}} k^\Zc \, (\delta\Thz)^2+\frac{1}{2} \, \sum_{\mathcal {C}} k^\Cc \, (\delta\Thc)^2,
\end{aligned}
\end{equation}
where $\Uc_\ell^{(s)}$ is the \textit{self-energy} (corresponding to the so-called \textit{cohesive energy} in the literature) and $\Uc_\ell^{(d)}$ is the \textit{dihedral energy}.
The analysis in a paper by Davini \cite{Davini_2014} applies here to the in-plane deformations, providing a continuum model of the graphene sheet within the framework of $\Gamma$-convergence theory. Hereafter, we concentrate on the out-of-plane deformations. 
With the notation introduced in Section~\ref{sec:STRAINS} we now write the bending energy more explicitly.
The self-energy can be written as
\begin{equation}\label{selfen1}
\mathcal{U}^{(s)}_\ell
= \sum_{\xb^\ell\in (L_1(\ell)\cup L_2(\ell))\cap\Omega}   \tau_0 \sum_{i = 
	1}^3\delta \vartheta_i^{(2)} (\xb^\ell),
\end{equation}
where $\sum_{i =1}^3\delta \vartheta_i^{(2)} (\xb^\ell)$ is given in \eqref{eq:sumdth_1} in terms of the out-of-plane component of the displacement $w$.
We further split the dihedral energy  $\Uc_\ell^{(d)}$ in
$$
\Uc_\ell^{(d)}:=\mathcal{U}^{\mathcal{Z}}_\ell+\mathcal{U}^{\mathcal{C}}_\ell,
$$
where
\begin{equation}\label{enZ}
\mathcal{U}^{\mathcal{Z}}_\ell=\frac{1}{2}\, k^{\mathcal{Z}} \sum_{\xb^\ell\in L_2(\ell)\cap\Omega}
\sum_{i=1}^3 \Bigg(\delta\Thz_{\pb_i\pb_{i+2}}(\xb^\ell)\Bigg)^2+ \Bigg(\delta\Thz_{\pb_i\pb_{i+1}}(\xb^\ell)\Bigg)^2
\end{equation}
is the contribution of the Z-dihedra, and
\begin{equation}\label{enC}
\mathcal{U}^{\mathcal{C}}_\ell=\frac{1}{2} k^{\mathcal{C}} \, \sum_{\xb^\ell\in L_2(\ell)\cap\Omega}
\sum_{i=1}^3 \Bigg(\delta\Thc_{\pb_i^+}(\xb^\ell)\Bigg)^2+ \Bigg(\delta\Thc_{\pb_i^-}(\xb^\ell)\Bigg)^2
\end{equation}
is the contribution of the C-dihedra. The Z- and C-dihedral angles appearing in \eqref{enZ} and \eqref{enC} are given in terms of $w$ in \eqref{C-wedge}-\eqref{CZ-wedge}.

In the next section we deduce, by means of a formal analysis, a continuous version of the discrete bending energy 
\begin{equation}\label{benen}
\mathcal{U}_\ell^{(b)}=\Uc_\ell^{(s)} +\mathcal{U}^{\mathcal{Z}}_\ell+\mathcal{U}^{\mathcal{C}}_\ell,
\end{equation}
from which we shall deduce
expressions for the sheet's bending stiffnesses. A rigorous analysis based on $\Gamma$-convergence theory will be done in a forthcoming paper \cite{Davini_2017}.

\section{The continuum limit}\label{SEC:continuum}

In this section we  find a continuous energy, defined over the domain $\Omega$, that 
approximates the discrete bending energy $\mathcal{U}_\ell^{(b)}$  defined over the lattice $(L_1(\ell)\cup L_2(\ell))\cap\Omega$. This is achieved by letting the lattice size $\ell$ go to zero so that
$(L_1(\ell)\cup L_2(\ell))\cap\Omega$ invades $\Omega$.
With this in mind, in place of a function $w:(L_1(\ell)\cup L_2(\ell))\cap\Omega\to \mathbb{R}$, we consider a twice continuously differentiable function $w:\Omega \to \mathbb{R}$.

\begin{figure}[h]
	\centering
	\includegraphics[scale=1]{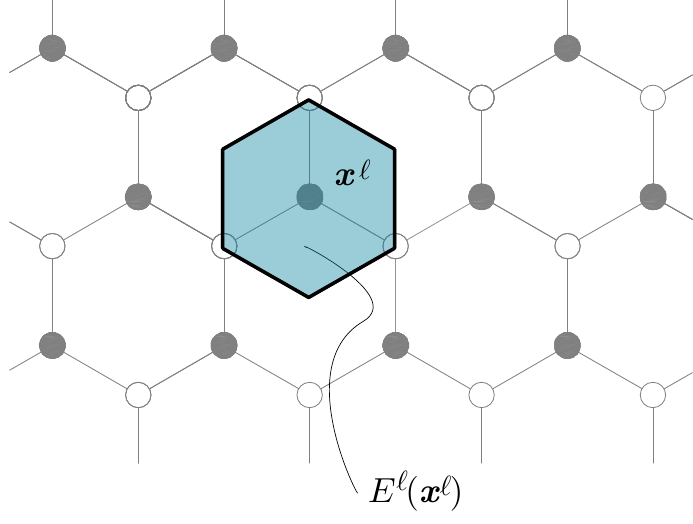}
	\caption{The hexagon $E^\ell(\xb^\ell)$.}
	\label{fig:hexagon}
\end{figure}

Given two vectors $\ab$ and $\bb$, with $\partial^2_{\ab\bb}w$ we denote the second partial derivative
of $w$ in the directions $\ab/|\ab|$ and $\bb/|\bb|$, that is
$$
\partial^2_{\ab\bb}w=\nabla^2 w \frac{\ab}{|\ab|}\cdot \frac{\bb}{|\bb|},
$$
where  $\nabla^2 w$ denotes the Hessian of $w$. Clearly, we also have
\begin{equation}\label{secder}
\partial^2_{\ab\bb}w(x_0)=\lim_{\ell\to 0}\frac{w(x_0+\ell \ab+\ell \bb)-w(x_0+\ell \bb)-w(x_0+\ell \ab)+w(x_0)}{\ell^2 |\ab| |\bb|}.
\end{equation}
The change of the Z-dihedra, see \eqref{CZ-wedge}$_2$, can be rewritten, after setting
\begin{equation}\label{ai}
\ab_i:=\pb_i-\pb_{i+2},
\end{equation}
as
\begin{align}
\delta\Thz_{\pb_i\pb_{i+2}}(\xb^\ell)&=\frac{2\sqrt{3}}{3\ell}[w(\xb^\ell+\ell\pb_i-\ell \pb_{i+2})-w(\xb^\ell)-w(\xb^\ell+\ell \pb_{i})+w(\xb^\ell+\ell \pb_{i+2})]\nonumber\\
&=\frac{2\sqrt{3}}{3\ell}[w(\xb^\ell+\ell\ab_i)-w(\xb^\ell)-w(\xb^\ell+\ell \ab_{i}+\ell \pb_{i+2})+w(\xb^\ell+\ell \pb_{i+2})]\nonumber\\
&=\frac{2\sqrt{3}}{3\ell} [- \partial^2_{\ab_i\pb_{i+2}}w(\xb^\ell) \ell^2 |\ab_i| |\pb_{i+2}|+o(\ell^2)]\nonumber\\
&=-2 \ell \partial^2_{\ab_i\pb_{i+2}}w(\xb^\ell) +o(\ell),\label{angZ1}
\end{align}
where the third equality follows from \eqref{secder}. Similarly, setting
\begin{equation}\label{bi}
\bb_i:=\pb_i-\pb_{i+1},
\end{equation}
we have that
\begin{align}
\delta\Thz_{\pb_i\pb_{i+1}}(\xb^\ell)&=
\frac{2\sqrt{3}}{3\ell}[w(\xb^\ell+\ell\pb_i-\ell \pb_{i+1})-w(\xb^\ell+\ell \pb_{i})+w(\xb^\ell+\ell \pb_{i+1})-w(\xb^\ell)]\nonumber\\
&=\frac{2\sqrt{3}}{3\ell}[w(\xb^\ell+\ell\bb_i)-w(\xb^\ell+\ell\bb_i+\ell \pb_{i+1})+w(\xb^\ell+\ell \pb_{i+1})-w(\xb^\ell)]\nonumber\\
&=-2 \ell \partial^2_{\bb_i\pb_{i+1}}w(\xb^\ell) +o(\ell).\label{angZ2}
\end{align}
Taking \eqref{eq: KINENER 3} into account, we may rewrite the vectors $\ab_i$ and $\bb_i$, defined in \eqref{ai} and \eqref{bi}, in terms of $\db_i$, for instance $\ab_1=\db_1$ and $\bb_1=-\db_3$,
and then rewrite the Z-dihedral energy, see \eqref{enZ} and rewritten below for the reader convenience, as
\begin{align}
\mathcal{U}^{\mathcal{Z}}_\ell&=\frac{1}{2}\, k^{\mathcal{Z}} \sum_{\xb^\ell\in L_2(\ell)\cap\Omega}
\sum_{i=1}^3 \Bigg(\delta\Thz_{\pb_i\pb_{i+2}}(\xb^\ell)\Bigg)^2+ \Bigg(\delta\Thz_{\pb_i\pb_{i+1}}(\xb^\ell)\Bigg)^2
\nonumber\\
&=\frac{1}{2} {4}\ell^2 k^{\mathcal{Z}} \sum_{\xb^\ell\in L_2(\ell)\cap\Omega} (\partial^2_{\db_1\pb_{3}}w(\xb^\ell))^2+ (\partial^2_{\db_1\pb_{1}}w(\xb^\ell))^2+(\partial^2_{\db_2\pb_{2}}w(\xb^\ell))^2\nonumber\\
&\hspace{3cm}+(\partial^2_{\db_2\pb_{3}}w(\xb^\ell))^2+(\partial^2_{\db_3\pb_{1}}w(\xb^\ell))^2+(\partial^2_{\db_3\pb_{2}}w(\xb^\ell))^2+o(\ell^2),\nonumber\\
&=\frac{1}{2} \frac{8\sqrt{3}}9 k^{\mathcal{Z}} \sum_{\xb^\ell\in L_2(\ell)\cap\Omega} \Bigg(\partial^2_{\db_1\pb_{3}}w(\xb^\ell))^2+ (\partial^2_{\db_1\pb_{1}}w(\xb^\ell))^2+(\partial^2_{\db_2\pb_{2}}w(\xb^\ell))^2\nonumber\\
&\hspace{2cm}+(\partial^2_{\db_2\pb_{3}}w(\xb^\ell))^2+(\partial^2_{\db_3\pb_{1}}w(\xb^\ell))^2+(\partial^2_{\db_3\pb_{2}}w(\xb^\ell))^2\Bigg)|E^\ell(\xb^\ell)|+o(\ell^2),\nonumber
\end{align}
where $|E^\ell(\xb^\ell)|=\ell^23\sqrt{3}/2$ is the area of the hexagon 
$E^\ell(\xb^\ell)$ of side $\ell$ centred at $\xb^\ell$ (see Fig. \ref{fig:hexagon}).

Let $\chi_{E^\ell(\xb^\ell)}(\xb)$ be the characteristic function of $E^\ell(\xb^\ell)$, i.e., it is equal to $1$ if $\xb\in  
E^\ell(\xb^\ell)$ and $0$ otherwise, and let 
\begin{align*}
W^\Zc_\ell(\xb):=\sum_{\xb^\ell\in L_2(\ell)\cap\Omega} &\Bigg((\partial^2_{\db_1\pb_{3}}w(\xb^\ell))^2+ (\partial^2_{\db_1\pb_{1}}w(\xb^\ell))^2+(\partial^2_{\db_2\pb_{2}}w(\xb^\ell))^2\nonumber\\
&+(\partial^2_{\db_2\pb_{3}}w(\xb^\ell))^2+(\partial^2_{\db_3\pb_{1}}w(\xb^\ell))^2+(\partial^2_{\db_3\pb_{2}}w(\xb^\ell))^2\Bigg)\chi_{E^\ell(\xb^\ell)}(\xb).
\end{align*}
Then, we may simply write
$$
\mathcal{U}^{\mathcal{Z}}_\ell=\frac{1}{2}\frac{8\sqrt{3}}9 k^{\mathcal{Z}}\int_\Omega W^\Zc_\ell(\xb)\,d\xb+o(\ell^2),
$$
and since $W^\Zc_\ell$ converges, as $\ell$ goes to zero, to
$$
(\partial^2_{\db_1\pb_{3}}w)^2+ (\partial^2_{\db_1\pb_{1}}w)^2+(\partial^2_{\db_2\pb_{2}}w)^2+(\partial^2_{\db_2\pb_{3}}w)^2+(\partial^2_{\db_3\pb_{1}}w)^2+(\partial^2_{\db_3\pb_{2}}w)^2,
$$
we deduce that
\begin{align}
\lim_{\ell\to 0}\mathcal{U}^{\mathcal{Z}}_\ell=\frac{1}{2}\frac{8\sqrt{3}}9 k^{\mathcal{Z}}\int_\Omega &(\partial^2_{\db_1\pb_{3}}w)^2+ (\partial^2_{\db_1\pb_{1}}w)^2
+(\partial^2_{\db_2\pb_{2}}w)^2\label{limenZ}\\
&+(\partial^2_{\db_2\pb_{3}}w)^2
+(\partial^2_{\db_3\pb_{1}}w)^2+(\partial^2_{\db_3\pb_{2}}w)^2\,d\xb=:\mathcal{U}^{\mathcal{Z}}_0(w).\nonumber
\end{align}
The functional $\mathcal{U}^{\mathcal{Z}}_0$, defined in \eqref{limenZ}, is the continuum limit of the Z-dihedral energy.

Working in a similar manner, we find the continuum limit of the C-dihedral energy:
\begin{equation}\label{limenC}
\lim_{\ell\to 0}\mathcal{U}^{\mathcal{C}}_\ell=\frac{1}{2} \frac{16\sqrt{3}}{9} k^{\mathcal{C}}\int_\Omega \sum_{i=1}^3 \Big(\partial^2_{\pb_i\pb_i^\perp}w\Big)^2\,d\xb
=:\mathcal{U}^{\mathcal{C}}_0(w),
\end{equation}
and the continuum limit of the self-energy:
\begin{align}
\lim_{\ell\to 0}\mathcal{U}^{(s)}_\ell=-\frac{4}9\tau_0 \int_\Omega ( \partial^2_{\pb_1 \pb_1}w+\partial^2_{\pb_2 \pb_2}w+\partial^2_{\pb_1 \pb_2}w\Big)^2\, d\xb
=:\mathcal{U}^{(s)}_0(w).\label{limentheta}
\end{align}
Detailed calculations leading to \eqref{limenC} and \eqref{limentheta} are found in Appendix \ref{Acont}.

The total bending limit energy is therefore
\begin{equation}\label{totalen}
\mathcal{U}^{(b)}_0(w):=\mathcal{U}^{\mathcal{Z}}_0(w)+\mathcal{U}^{\mathcal{C}}_0(w)+\mathcal{U}^{(s)}_0(w). 
\end{equation}

\section{The equivalent plate equation}\label{plate}
In this section we rewrite the limit energies in a more amenable form.
We start by manipulating the limit C-dihedral energy. We first note that
\begin{align}
\partial^2_{\pb_2\pb_2^\perp}w&=\nabla^2 w \pb_2\cdot\pb_2^\perp=\nabla^2 w (-\frac 12 \pb_1+\frac{\sqrt{3}}2 \pb_1^\perp)\cdot (-\frac{\sqrt{3}}2 \pb_1-\frac 12\pb_1^\perp)\nonumber\\
&=\frac{\sqrt{3}}4\partial^2_{\pb_1\pb_1}w-\frac{\sqrt{3}}4\partial^2_{\pb_1^\perp\pb_1^\perp}w
-\frac{1}2\partial^2_{\pb_1\pb_1^\perp}w,\label{calcpar}
\end{align}
and similarly
$$
\partial^2_{\pb_3\pb_3^\perp}w=
-\frac{\sqrt{3}}4\partial^2_{\pb_1\pb_1}w+\frac{\sqrt{3}}4\partial^2_{\pb_1^\perp\pb_1^\perp}w
-\frac{1}2\partial^2_{\pb_1\pb_1^\perp}w,
$$
from which we find that 
\begin{align}
\sum_{i=1}^3 \Big(\partial^2_{\pb_i\pb_i^\perp}w\Big)^2&=(\partial^2_{\pb_1\pb_1^\perp}w)^2+
\Big(\frac{\sqrt{3}}4\partial^2_{\pb_1\pb_1}w-\frac{\sqrt{3}}4\partial^2_{\pb_1^\perp\pb_1^\perp}w
-\frac{1}2\partial^2_{\pb_1\pb_1^\perp}w\Big)^2\nonumber\\
&\quad+
\Big(-\frac{\sqrt{3}}4\partial^2_{\pb_1\pb_1}w+\frac{\sqrt{3}}4\partial^2_{\pb_1^\perp\pb_1^\perp}w
-\frac{1}2\partial^2_{\pb_1\pb_1^\perp}w)^2\nonumber\\
&=\frac 32 (\partial^2_{\pb_1\pb_1^\perp}w)^2+ \frac 38
\Big(\partial^2_{\pb_1\pb_1}w-\partial^2_{\pb_1^\perp\pb_1^\perp}w\Big)^2\nonumber\\
&=\frac 32 (\partial^2_{\pb_1\pb_1^\perp}w)^2+ \frac 38
\Big(\partial^2_{\pb_1\pb_1}w+\partial^2_{\pb_1^\perp\pb_1^\perp}w\Big)^2-\frac 32 \partial^2_{\pb_1\pb_1}w\partial^2_{\pb_1^\perp\pb_1^\perp}w\nonumber\\
&=\frac 38 (\Delta w)^2- \frac 32 \det \nabla^2 w,\label{sumders}
\end{align}
where $\Delta w$ denotes the Laplacian of $w$.
Hence, the C-dihedral energy defined in \eqref{limenC} rewrites as
\begin{align}
\mathcal{U}^{\mathcal{C}}_0(w)&=\frac{1}{2} \frac{16\sqrt{3}}{9} k^{\mathcal{C}}\int_\Omega \sum_{i=1}^3 \Big(\partial^2_{\pb_i\pb_i^\perp}w\Big)^2\,d\xb=
\frac{1}{2} \frac{2\sqrt{3}}{3} k^{\mathcal{C}}\int_\Omega (\Delta w)^2- 4 \det \nabla^2 w\,d\xb.
\label{enC}
\end{align}

We now tackle the Z-dihedral energy. Recalling \eqref{limenZ}, by a calculation similar to that carried on in \eqref{calcpar} we find
$$
\partial^2_{\db_1\pb_3}w=
-\frac{\sqrt{3}}2\partial^2_{\db_1\db_1}w-\frac{1}2\partial^2_{\db_1\db_1^\perp}w \quad \mbox{and}\quad
\partial^2_{\db_1\pb_1}w=
+\frac{\sqrt{3}}2\partial^2_{\db_1\db_1}w-\frac{1}2\partial^2_{\db_1\db_1^\perp}w,
$$
where $\db_i^\perp=\eb_3\times \db_i$, and from these equations we deduce that
$$
(\partial^2_{\db_1\pb_1}w)^2+(\partial^2_{\db_1\pb_3}w)^2=\frac 32 (\partial^2_{\db_1\db_1}w)^2+
\frac 12 (\partial^2_{\db_1\db_1^\perp}w)^2.
$$
Similar identities hold for $\db_2$ and $\db_3$. Thence, we find that  the Z-dihedral energy
takes the form
\begin{align*}
\mathcal{U}^{\mathcal{Z}}_0(w)&=\frac{1}{2}\frac{8\sqrt{3}}9 k^{\mathcal{Z}}\int_\Omega (\partial^2_{\db_1\pb_{3}}w)^2+ (\partial^2_{\db_1\pb_{1}}w)^2
+(\partial^2_{\db_2\pb_{2}}w)^2\\
&\hspace{3cm}+(\partial^2_{\db_2\pb_{3}}w)^2
+(\partial^2_{\db_3\pb_{1}}w)^2+(\partial^2_{\db_3\pb_{2}}w)^2\,d\xb\\
&=\frac{1}{2}\frac{8\sqrt{3}}9 k^{\mathcal{Z}}\int_\Omega \frac 32 \sum_{i=1}^3(\partial^2_{\db_i\db_i}w)^2+
\frac 12 \sum_{i=1}^3(\partial^2_{\db_i\db_i^\perp}w)^2 \,d\xb.
\end{align*}
The second sum is equal, as it can be checked, to the last line of \eqref{sumders}, and with a similar calculation we also find that
$$
\sum_{i=1}^3(\partial^2_{\db_i\db_i}w)^2=\frac 98 (\Delta w)^2- \frac 32 \det \nabla^2 w,
$$
and hence
\begin{align}
\mathcal{U}^{\mathcal{Z}}_0(w)
&=\frac{1}{2}\frac{8\sqrt{3}}9 k^{\mathcal{Z}}\int_\Omega \frac 32\Bigg(\frac 98 (\Delta w)^2- \frac 32 \det \nabla^2 w\Bigg)+
\frac 12 \Bigg(\frac 38 (\Delta w)^2- \frac 32 \det \nabla^2 w\Bigg) \,d\xb
\nonumber\\
&=\frac{1}{2}\frac{8\sqrt{3}}9 k^{\mathcal{Z}}\int_\Omega \frac{15}8 (\Delta w)^2- 3 \det \nabla^2 w\,d\xb
\nonumber\\
&=\frac{1}{2}\frac{5\sqrt{3}}3 k^{\mathcal{Z}}\int_\Omega  (\Delta w)^2- \frac{8}5 \det \nabla^2 w\,d\xb.
\label{enZ}
\end{align}

We now deal with the self-energy. Again with a calculation similar to that carried on in \eqref{calcpar} we find
\begin{align*}
\partial^2_{\pb_1\pb_2}w
&=-\frac{1}2\partial^2_{\pb_1\pb_1}w+\frac{\sqrt{3}}2\partial^2_{\pb_1\pb_1^\perp}w
\\
\partial^2_{\pb_2\pb_2}w&=\frac{1}4\partial^2_{\pb_1\pb_1}w+\frac{{3}}4\partial^2_{\pb_1^\perp\pb_1^\perp}w
-\frac{\sqrt{3}}2\partial^2_{\pb_1\pb_1^\perp}w,
\end{align*}
and hence
\begin{align}
\mathcal{U}^{(s)}_0(w)&=-\frac{4}9\tau_0 \int_\Omega ( \partial^2_{\pb_1 \pb_1}w+\partial^2_{\pb_2 \pb_2}w+\partial^2_{\pb_1 \pb_2}w\Big)^2\, d\xb\nonumber\\
&=-\frac{4}9\tau_0 \int_\Omega ( \frac 34 \partial^2_{\pb_1 \pb_1}w+ \frac 34 \partial^2_{\pb_1^\perp\pb_1^\perp}w\Big)^2\, d\xb\nonumber\\
&=-\frac 12 \frac{\tau_0}2 \int_\Omega ( \Delta w)^2\, d\xb.\label{enth}
\end{align}

By summing \eqref{enC}, \eqref{enZ}, and \eqref{enth}, we find the total bending energy, defined in \eqref{totalen}:
\begin{align}\label{entot}
\mathcal{U}^{(b)}_0(w)
=&\frac 12 \int_\Omega\Bigg(\frac{5\sqrt{3}}3 k^{\mathcal{Z}}+\frac{2\sqrt{3}}{3} k^{\mathcal{C}}-\frac{\tau_0}2\Bigg)( \Delta w)^2\nonumber\\
&\hspace{3cm}+\Bigg(- \frac{8}5\frac{5\sqrt{3}}3 k^{\mathcal{Z}}-4\frac{2\sqrt{3}}{3} k^{\mathcal{C}}\Bigg)\det \nabla^2 w\,d\xb,\\
&=\frac 12 \int_\Omega \Dc ( \Delta w)^2+\Dc_G\det \nabla^2 w\,d\xb,
\end{align}
where 
\begin{equation}\label{stiffnesses}
\Dc:= \frac{5\sqrt{3}}3 k^{\mathcal{Z}}+\frac{2\sqrt{3}}{3} k^{\mathcal{C}}-\frac{\tau_0}2, \qquad 
\Dc_G:=- \frac{8}5\frac{5\sqrt{3}}3 k^{\mathcal{Z}}-4\frac{2\sqrt{3}}{3} k^{\mathcal{C}}.
\end{equation}
are the \textit{bending} and the \textit{Gaussian stiffnesses}, respectively. $\Dc_G$ is called Gaussian because it multiplies the \textit{Gaussian curvature} $\det\nabla^2w$, while $\Delta w$ is twice the \textit{mean curvature}. The analytical expression for the bending stiffness \eqref{stiffnesses}$_1$ coincides with the one deduced in \cite{Lu_2009, Favata_2016b}, within a discrete mechanical framework, if one assumes that $k^{\mathcal{Z}}\equiv k^{\mathcal{C}}$. It clearly shows that the origin of the bending stiffness is twofold: a part depends on the dihedral contribution, and a part on the self-stress. Quite surprisingly, the \textit{self-stress has no role in the Gaussian stiffness}. It is worth noticing that in the above approach there is no need of introducing any questionable effective \textit{thickness} parameter.

\section{Numerical results}\label{num}
In this section, we adopt the 2nd-generation Brenner potential \cite{Brenner_2002} to obtain quantitative results for the continuum material parameters deduced in Sec. \ref{plate}.

The 2nd-generation REBO potentials developed for hydrocarbons by Brenner \emph{et al.} in \cite{Brenner_2002} accommodate  up to third-nearest-neighbor interactions through a bond-order function depending, in particular, on dihedral angles.  Following Appendix B of \cite{Favata_2016}, we here give a short account of the form of this potential.

The  binding energy $V$ of an atomic aggregate is given as a sum over nearest neighbors:
\begin{equation}\label{V}
V=\sum_i\sum_{j<i} V_{ij}\,;
\end{equation}
the interatomic potential $V_{ij}$ is given by 
\begin{equation}\label{Vij}
V_{ij}=V_R(r_{ij})+b_{ij}V_A(r_{ij}),
\end{equation}
where the individual effects of the \emph{repulsion} and \emph{attraction functions} $V_R(r_{ij})$ and $V_A(r_{ij})$, which model pair-wise interactions of  atoms $i$ and $j$ depending on their distance $r_{ij}$, are modulated by the \emph{bond-order function} $b_{ij}$. The repulsion and attraction functions have the following forms:
\begin{equation}\label{VA}
\begin{aligned}
V_A(r)&=-f^C(r)\sum_{n=1}^{3}B_n e^{-\beta_n r}\,,\\
V_R(r)&=f^C(r)\left( 1 + \frac{Q}{r} \right) A e^{-\alpha r}\,,
\end{aligned}
\end{equation}
where $f^C(r)$ is a \emph{cutoff function} limiting the range of covalent interactions, and where $Q$, $A$, $B_n$, $\alpha$, and $\beta$, are parameters to be chosen fit to some material-specific dataset. The remaining ingredient in \eqref{Vij} is the \emph{bond-order function}:
\begin{equation}\label{bij}
b_{ij}=\frac{1}{2}(b_{ij}^{\sigma-\pi}+b_{ji}^{\sigma-\pi})+b_{ij}^\pi\,,
\end{equation}
where apexes $\sigma$ and $\pi$ refer to two types of bonds: the strong covalent $\sigma$-bonds between atoms in one and the same given plane, and the $\pi$-bonds responsible for interlayer interactions, which are perpendicular to the plane of $\sigma$-bonds. 
The role of function $b_{ij}^{\sigma-\pi}$ is to account for the local coordination of, and the bond angles relative to, atoms $i$ and $j$; its form is:
\begin{equation}\label{G}
b_{ij}^{\sigma-\pi}=\left(1+\sum_{k\neq i,j} f_{ik}^C(r_{ik})G(\cos\theta_{ijk}) \,e^{\lambda_{ijk}}+P_{ij}(N_i^C,N_i^H)  \right)^{-1/2}\,.
\end{equation}
Here, for each fixed pair of indices $(i,j)$, (a) the cutoff function $f_{ik}^C$ limits the interactions of atom $i$ to those with its nearest neighbors; (b) $\lambda_{ijk}$ is a string of parameters designed to prevent attraction in some specific situations; (c) function $P_{ij}$ depends on $N_i^C$ and $N_i^H$, the numbers of $C$ and $H$ atoms that are nearest neighbors of atom $i$; it is meant to adjust the bond-order function 	according to the environment of the C atoms in one or another molecule; (d) for solid-state carbon, the values of  both the string $\lambda_{ijk}$ and the function $P_{ij}$ are taken null; (e)  function $G$ modulates the contribution of each nearest neighbour of atom $i$ in terms of the cosine of the angle between the $ij$ and $ik$ bonds; its analytic form is given by three sixth-order polynomial splines.	
Function $b_{ij}^\pi$ is given a split representation:
\begin{equation}
b_{ij}^\pi=\Pi_{ij}^{RC}+b_{ij}^{DH},
\end{equation}
where the first addendum $\Pi_{ij}^{RC}$ depends on whether the bond between atoms $i$ and $j$ has a radical character and on whether it is part of a conjugated system, while the second addendum $b_{ij}^{DH}$ depends on dihedral angles and has the following form:
\begin{equation}
b_{ij}^{DH}=T_{ij}(N_i^t,N_j^t,N_{ij}^{\rm conj})\left(\sum_{k(\neq i,j)}\sum_{k(\neq i,j)}\big( 1-\cos^2\Theta_{ijkl} \big)f_{ik}^C(r_{ik})f_{jl}^C(r_{jl})  \right)\,,
\end{equation}
where function $T_{ij}$ is a tricubic spline depending on $N_i^t=N_i^C+N_i^H$, $N_j^t$, and $N_{ij}^{\rm conj}$, a function  of local conjugation, and the dihedral angle is defined as
\begin{equation}
\cos\Theta_{ijkl}=\nb_{jik}\cdot\nb_{ijl}, \quad \nb_{jik}=\frac{\rb_{ji}\times\rb_{ik}}{|\rb_{ji}\times\rb_{ik}|}, \; \nb_{jil}=\frac{\rb_{ij}\times\rb_{il}}{|\rb_{ij}\times\rb_{jl}|}.
\end{equation}

The values of the constant $k^\Zc$ and $k^\Cc$ can be deduced by deriving twice of the potential, and computing the result in the ground state (GS): $r_{ij}=\ell$, $\theta_{ijk}=2/3\pi$, $\Theta_{ijkl}=0$. In particular, we find:
\begin{equation}
k^\Theta:=k^\Zc=k^\Cc=\partial^2_{\Theta_{ijkl}}V_{ij}|_{GS}=2TV_A(\ell),
\end{equation}
where $T$ is the value of $T_{ij}$ in the GS. 

\begin{remark}
	With this notation the bending stiffness becomes:
	$$
	\Dc=\frac{7\sqrt{3}}{3}k^\Theta-\frac{\tau_0}{2}.
	$$
	This expression coincides with that given in [28]:
	$$
	\mathcal{D}=\frac{V_A(r_0)}{2}\left(  (b_0^{\sigma-\pi})'-\frac{14 T_0}{\sqrt{3}}\right),
	$$
	after noticing that 
	$$
	V_A(r_0)(b_0^{\sigma-\pi})'\equiv-\tau_0, \quad -V_A(r_0)\left(  \frac{7 T_0}{\sqrt{3}}\right)\equiv 2TV_A(\ell) \frac{7}{\sqrt{3}}= \frac{7}{\sqrt{3}}k^\Theta.
	$$
	
	Instead, in references [4] and [21] the dihedral energies are not contemplated and the bending stiffness found, up to notational differences, coincides with   ours after setting $k^\Theta=0$.
\end{remark}

With the values reported in \cite{Brenner_2002}, we get:
\begin{equation}\label{kTh}
k^\Theta=0.0282\;\textrm{nN nm}=0.1764 \;\textrm{eV}.
\end{equation}
From \cite{Favata_2016}, we take the value of the selfstress $\tau_0$:
\begin{equation}\label{tau0}
\tau_0=-0.2209\;\textrm{nN nm}=-1.3787\;\textrm{eV}.
\end{equation}
With \eqref{kTh} and \eqref{tau0}, we obtain:
\begin{equation}\label{rigid}
\begin{aligned}
& \Dc=\frac{7\sqrt{3}}{3}k^\Theta-\frac{\tau_0}{2}=0.2247 \;\textrm{nN nm}=1.4022 \;\textrm{eV},\\
& \Dc_G=-16\frac{\sqrt{3}}{3}k^\Theta=-0.2610\;\textrm{nN nm}=-1.6293\;\textrm{eV}.
\end{aligned}
\end{equation}
The value  of $\Dc$ is in complete agreement with the literature \cite{LuJ1997,Wei_2013}; from \eqref{kTh} and \eqref{tau0} it is possible to check that  the contribution of the self-stress and the dihedral stiffness amounts to about  $49.16$\% and $50.84$\% of the total, respectively. Neither analytical evaluation of $\Dc_G$, nor MD computations, have been proposed so far. The value we obtain is in good agreement with the value of $-1.52$ eV, reported in \cite{Wei_2013} and  determined by means of DFT.

\section{Conclusions}

Starting form a discrete model inferred from MD, we have deduced a continuum theory describing the \textit{bending behavior} of a graphene sheet. Atomic interactions have been modeled by exploiting the main features of the 2nd-generation Brenner potential and adopting a quadratic approximation of the energy. The deduced continuum limit   fully describes the bending behavior of graphene. To our knowledge, it is the first time that an analytical expression of the \textit{Gaussian stiffness} is given  and an explanation of its origins at the atomistic scale is provided. We also derived a quantitative evaluation of the related constitutive parameters.

\begin{footnotesize}
	\section*{Acknowledgments}	
	AF acknowledges the financial support of Sapienza University of Rome (Progetto d'Ateneo 2016 --- ``Multiscale Mechanics of 2D Materials: Modeling and Applications''). 
	
\end{footnotesize}

\appendix
\section{Appendix}\label{appendix}
In order to compute the strain measures for small changes of configuration of the graphene foil, we write the displacements of the nodes in the form
$${\bf u} (\xb^\ell)= \xi\ufb(\xb^\ell),$$
where $\xi$ is a positive scalar measuring smallness and $\ufb:=\frac{\bf u}{\xi}$ stands for the displacement distribution normalized accordingly.

\subsection{Change of the bond angle}\label{app:WANGLE}

Let us define the bond angle as
\begin{equation}\label{eq:wangle}
\cos(\vartheta_i(\xi))=\left(\frac{\mathbf{m}(\xi)\cdot\mathbf{n}(\xi)}{|\mathbf{m}(\xi)||\mathbf{n}(\xi)|}\right) ,
\end{equation} 
with
$$\mathbf{m}(\xi):=\ell\pb_{i+1}+\xi (\ufb_{i+1} - \ufb_0)\quad \mbox{and} \quad \mathbf{n}(\xi):=\ell\pb_{i+2}+\xi (\ufb_{i+2} - \ufb_0),$$
where we have set
\begin{equation}
\ufb_0:=\ufb(\xb^\ell), \quad \ufb_{i}:=\ufb(\xb^\ell+\ell\pb_i).
\end{equation}
Then, from Taylor's expansion we get
\begin{equation}\label{eq:wangle_1}
\vartheta_i(\xi)= \vartheta_i(0)+\vartheta_i^\prime(0)\,\xi+\frac{1}{2}\,\vartheta_i^{\prime \prime}(0) \xi^2 +O(\xi^3),
\end{equation}
where the various terms can be calculated by successive differentiations of Eq.~\eqref{eq:wangle}. Thus,
\begin{equation}
\begin{aligned}
-\sin(\vartheta_i(0))\,&\vartheta_i^\prime(0)=\left.\left(\frac{\mathbf{m}(\xi)\cdot\mathbf{n}(\xi)}{|\mathbf{m}(\xi)||\mathbf{n}(\xi)|}\right)^\prime\right|_{\xi=0}\\
&=\frac{\mathbf{m}'(\xi)\cdot\mathbf{n}(\xi)+\mathbf{m}(\xi)\cdot\mathbf{n}'(\xi)}{|\mathbf{m}(\xi)||\mathbf{n}(\xi)|}-\frac{\mathbf{m}(\xi)\cdot\mathbf{n}(\xi)}{|\mathbf{m}(\xi)||\mathbf{n}(\xi)|}\left.\left(\frac{\mathbf{m}(\xi)\cdot\mathbf{m}'(\xi)}{|\mathbf{m}(\xi)|^2} +\frac{\mathbf{n}(\xi)\cdot\mathbf{n}'(\xi)}{|\mathbf{n}(\xi)|^2} \right)\right|_{\xi=0},
\end{aligned}
\end{equation}
which yields
\begin{equation}\label{eq:wangle1st}
\vartheta_i^\prime(0)=-\frac{\pb_{i+2} +\frac{1}{2}\pb_{i+1}}{|\pb_{i+2} +\frac{1}{2}\pb_{i+1}|}\cdot(\ufb_{i+1} - \ufb_0) - \frac{\pb_{i+1} +\frac{1}{2}\pb_{i+2}}{|\pb_{i+1} +\frac{1}{2}\pb_{i+2}|}\cdot(\ufb_{i+2} - \ufb_0),
\end{equation}
where we take into account that $\sin(\vartheta_i(0))=\frac{\sqrt{3}}{2}$, $|\mathbf{m}(0)|=|\mathbf{n}(0)|=\ell$, $|\pb_{i+1} +\frac{1}{2}\pb_{i+2}|=|\pb_{i+2} +\frac{1}{2}\pb_{i+1}| = \sqrt{3}/2$ and $\pb_i\cdot\pb_{i+1}=-1/2$.

Moreover, by differentiating Eq.~\eqref{eq:wangle} twice, we get
\begin{equation}
-\cos(\vartheta_i(0))\,\vartheta^\prime_i(0)^2-\sin(\vartheta_i(0))\,\vartheta_i^{\prime\prime}(0)=\left. \left(\frac{\mathbf{m}(\xi)\cdot\mathbf{n}(\xi)}{|\mathbf{m}(\xi)||\mathbf{n}(\xi)|}\right)^{\prime\prime}\right |_{\xi=0},
\end{equation}
which gives
\begin{equation}\label{eq:theta(0)''}
\vartheta_i^{\prime\prime}(0)= -\frac{1}{\sin\vartheta_i(0)}\left(\cos\vartheta_i(0)\vartheta^\prime_i(0)^2+\left.\left(\frac{\mathbf{m}(\xi)\cdot\mathbf{n}(\xi)}{|\mathbf{m}(\xi)||\mathbf{n}(\xi)|}\right)''\right|_{\xi=0}  \right).
\end{equation} 
Computations yield that
\begin{equation}\label{second}
\begin{aligned}
\left(\frac{\mathbf{m}\cdot\mathbf{n}}{|\mathbf{m}||\mathbf{n}|}\right)''&=\frac{\mathbf{m}''}{|\mathbf{m}|}\cdot\frac{\mathbf{n}}{|\mathbf{n}|}+2\,\frac{\mathbf{m}'}{|\mathbf{m}|}\cdot\frac{\mathbf{n}'}{|\mathbf{n}|}+\frac{\mathbf{m}}{|\mathbf{m}|}\cdot\frac{\mathbf{n}''}{|\mathbf{n}|}\\
&-\left[\frac{\mathbf{m}'}{|\mathbf{m}|}\cdot\frac{\mathbf{n}}{|\mathbf{n}|}+\frac{\mathbf{m}}{|\mathbf{m}|}\cdot\frac{\mathbf{n}'}{|\mathbf{n}|}+\left(\frac{\mathbf{m}}{|\mathbf{m}|}\cdot\frac{\mathbf{n}}{|\mathbf{n}|}\right)'  \right] \left(\frac{\mathbf{m}'}{|\mathbf{m}|}\cdot\frac{\mathbf{m}}{|\mathbf{m}|}+\frac{\mathbf{n}'}{|\mathbf{n}|}\cdot\frac{\mathbf{n}}{|\mathbf{n}|}  \right)\\
&-\frac{\mathbf{m}}{|\mathbf{m}|}\cdot\frac{\mathbf{n}}{|\mathbf{n}|}\left[ \left(\frac{|\mathbf{m}'|}{|\mathbf{m}|}\right)^2+\frac{\mathbf{m}}{|\mathbf{m}|}\cdot\frac{\mathbf{m}''}{|\mathbf{m}|}-2\left(\frac{\mathbf{m}'}{|\mathbf{m}|}\cdot\frac{\mathbf{m}}{|\mathbf{m}|} \right)^2\right.\\  
& \left. +
\left(\frac{|\mathbf{n}'|}{|\mathbf{n}|}\right)^2+\frac{\mathbf{n}}{|\mathbf{n}|}\cdot\frac{\mathbf{n}''}{|\mathbf{n}|}-2\left(\frac{\mathbf{n}'}{|\mathbf{n}|}\cdot\frac{\mathbf{n}}{|\mathbf{n}|} \right)^2
\right]\,.
\end{aligned}
\end{equation}
Since $\mathbf{m}''(0)=\mathbf{n}''(0)=\mathbf{0}$ and $\left(\frac{\mathbf{m}\cdot\mathbf{n}}{|\mathbf{m}||\mathbf{n}|}\right)'\Big|_{\xi=0}=-\sin\vartheta_i(0)\vartheta_i^\prime(0)$, we finally have that
\begin{equation}\label{eq:theta(0)''_1}
\begin{aligned}
\left.\left(\frac{\mathbf{m}\cdot\mathbf{n}}{|\mathbf{m}||\mathbf{n}|}\right)''\right|_{\xi=0}
&=\frac{2}{\ell^2}(\ufb_{i+1} - \ufb_0)\cdot(\ufb_{i+2} - \ufb_0)-\left(\frac{(\ufb_{i+1} - \ufb_0)}{\ell}\cdot\pb_{i+1}+\frac{(\ufb_{i+2} - \ufb_0)}{\ell}\cdot\pb_{i+2}  \right)\times\\
&\times\left(\frac{(\ufb_{i+1} - \ufb_0)}{\ell}\cdot\pb_{i+2}+\frac{(\ufb_{i+2} - \ufb_0)}{\ell}\cdot\pb_{i+1} -\sin\vartheta_i(0)\vartheta_i^\prime(0)  \right)\\
& -(\pb_{i+1}\cdot\pb_{i+2})\left(\frac{1}{\ell^2}|\ufb_{i+1} - \ufb_0|^2-2\frac{1}{\ell^2}\Big( \pb_{i+1}\cdot(\ufb_{i+1} - \ufb_0) \Big)^2
+\frac{1}{\ell^2}|\ufb_{i+2} - \ufb_0|^2 \right.\\ 
&\left.-2\frac{1}{\ell^2}\Big( \pb_{i+2}\cdot(\ufb_{i+2} - \ufb_0) \Big)^2 
\right).
\end{aligned}
\end{equation}

\vspace{1cm}

All in all,  we have that:
\begin{equation}
\begin{aligned}
& \vartheta_i''(0)=-\frac{2}{\sqrt{3}}\Bigg[-\frac{1}{2}\vartheta_i'(0){ ^2}+\frac{2}{\ell^2}(\ufb_{i+1} - \ufb_0)\cdot(\ufb_{i+2} - \ufb_0)-\\
&\left(\frac{(\ufb_{i+1} - \ufb_0)}{\ell}\cdot\pb_{i+1}+\frac{(\ufb_{i+2} - \ufb_0)}{\ell}\cdot\pb_{i+2}  \right)\times\\
&\times\left(\frac{(\ufb_{i+1} - \ufb_0)}{\ell}\cdot\pb_{i+2}+\frac{(\ufb_{i+2} - \mathfrak{u}_0)}{\ell}\cdot\pb_{i+1} -\frac{\sqrt{3}}{2}\vartheta_i^\prime(0)  \right)\\
& -\frac{\pb_{i+1}\cdot\pb_{i+2}}{\ell^2}\Big(|\ufb_{i+1} - \ufb_0|^2-2\big( \pb_{i+1}\cdot(\ufb_{i+1} - \ufb_0) \big)^2
+|\ufb_{i+2} - \ufb_0|^2 \\ 
&-2\big( \pb_{i+2}\cdot(\ufb_{i+2} - \ufb_0) \big)^2    \Bigg]\,.
\end{aligned}
\end{equation}
Recalling that $\delta \vartheta_i=\vartheta_i^\prime(0)\,\xi+\frac{1}{2}\,\vartheta_i^{\prime \prime}(0) \xi^2 +O(\xi^3)$ and that $\mathbf{u}=\xi \ufb$, we get:
\begin{equation}\label{eq:deltatheta}
\delta\vartheta_i=\delta\vartheta_i^{(1)}+\delta\vartheta_i^{(2)}+O(\xi^3),
\end{equation}
with
\begin{equation}\label{eq:vtheta'0}
\delta\vartheta_i^{(1)}=\vartheta_i^\prime(0)\xi=-\frac{1}{\ell} \frac{\pb_{i+2} +\frac{1}{2}\pb_{i+1}}{|\pb_{i+2} +\frac{1}{2}\pb_{i+1}|}\cdot(\mathbf{u}_{i+1} - \mathbf{u}_0) - \frac{1}{\ell} \frac{\pb_{i+1} +\frac{1}{2}\pb_{i+2}}{|\pb_{i+1} +\frac{1}{2}\pb_{i+2}|}\cdot(\mathbf{u}_{i+2} - \mathbf{u}_0).
\end{equation}
and
\begin{equation}
\begin{aligned}
&\delta\vartheta_i^{(2)}=\frac{1}{2}\vartheta_i^{\prime\prime}(0)\xi^2= -\frac{1}{\sqrt{3}}\Bigg[-\frac{1}{2}{(\xi\vartheta_i'(0)) ^2}+\frac{2}{\ell^2}(\mathbf{u}_{i+1} - \mathbf{u}_0)\cdot(\mathbf{u}_{i+2} - \mathbf{u}_0)-\\
&\left(\frac{(\mathbf{u}_{i+1} - \mathbf{u}_0)}{\ell^2}\cdot\pb_{i+1}+\frac{(\mathbf{u}_{i+2} - \mathbf{u}_0)}{\ell^2}\cdot\pb_{i+2}  \right)\times\\
&\times\left(\frac{(\mathbf{u}_{i+1} - \mathbf{u}_0)}{\ell^2}\cdot\pb_{i+2}+\frac{(\mathbf{u}_{i+2} - \mathbf{u}_0)}{\ell^2}\cdot\pb_{i+1} -\frac{\sqrt{3}}{2}{\xi}\vartheta_i^\prime(0)  \right)
\\
& -\frac{\pb_{i+1}\cdot\pb_{i+2}}{\ell^4}\Big(|\mathbf{u}_{i+1} - \mathbf{u}_0|^2-2\frac{1}{\ell^2}\big( \pb_{i+1}\cdot(\mathbf{u}_{i+1} - \mathbf{u}_0) \big)^2
+|\mathbf{u}_{i+2} - \mathbf{u}_0|^2 \\ 
&-2\frac{1}{\ell^2}\big( \pb_{i+2}\cdot(\mathbf{u}_{i+2} - \mathbf{u}_0) \big)^2    \Bigg].
\end{aligned}
\end{equation}
We write \eqref{eq:vtheta'0} in the simpler form
\begin{equation}\label{eq:vtheta'0_2}
\delta\vartheta_i^{(1)}=-\frac{1}{\ell} (\mathbf{u}_{i+1} - \mathbf{u}_0)\cdot\pb^\perp_{i+1} + \frac{1}{\ell} (\mathbf{u}_{i+2} - \mathbf{u}_0)\cdot \pb^\perp_{i+2},
\end{equation}
where the unit vectors $\pb^\perp_{i+1}$'s are defined by \eqref{piperp}. 

\subsection{Change of the dihedral  angle}\label{app:DANGLE}
\begin{figure}[h]
	\centering
	\includegraphics[scale=1]{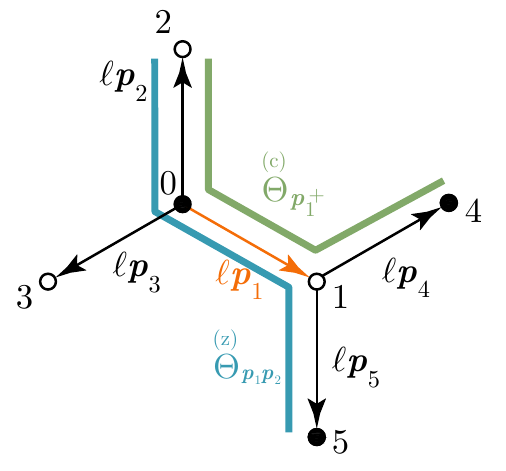}
	\caption{C-dihedral angle $\Thc_{\pb_1^+}$ (green) and Z-dihedral angle $\Thz_{\pb_1\pb_2}$ (blue). }
	\label{fig:cell_app}
\end{figure}

To fix the ideas, we focus on the  dihedral angles $\delta\Thc_{\pb_1^+}$, and  $\delta\Thz_{\pb_1\pb_2}$, sketched in Fig. \ref{fig:cell_app}; the other strains can be obtained in analogous manner.  

The first order approximation of the dihedral angle is all we need to evaluate the corresponding energy contribution. 

Let us denote by 
\begin{equation}
\qb_1=\ell\pb_1+(\ufb_1-\ufb_0)\xi, \quad \qb_2=\ell\pb_2+(\ufb_2-\ufb_0)\xi, \quad \qb_4=\ell\pb_4+(\ufb_4-\ufb_1)\xi
\end{equation}
the edge vectors after the deformation. We have that
\begin{equation}\label{Thc1+}
|\sin\Big( \Thc_{\pb_1^+}(\xi)\Big)|=\frac{|(\qb_1\times\qb_2)\times(\qb_1\times\qb_4)|}{|\qb_1\times\qb_2||\qb_1\times\qb_4|};
\end{equation}
it is easy to see that
\begin{equation}
\begin{aligned}
& \qb_1\times\qb_2=\frac{\sqrt{3}}{2}\ell^2\eb_3+ \Big( \ell\pb_1\times(\ufb_2-\ufb_0) -\ell\pb_2\times(\ufb_1-\ufb_0)\Big)\xi+O(\xi^2),\\
&\qb_1\times\qb_4=\frac{\sqrt{3}}{2}\ell^2\eb_3+\Big( \ell\pb_1\times(\ufb_4-\ufb_1) -\ell\pb_4\times(\ufb_1-\ufb_0)\Big)\xi+O(\xi^2),
\end{aligned}
\end{equation}
whence
\begin{equation}
\begin{aligned}
(\qb_1\times\qb_2)\times(\qb_1\times\qb_4)=\frac{\sqrt{3}}{2}\eb_3&\times\Big( \ell\pb_1\times(\ufb_4-\ufb_1)-\ell\pb_4\times(\ufb_1-\ufb_0) +\\
&+\ell\pb_2\times(\ufb_1-\ufb_0)-\ell\pb_1\times(\ufb_2-\ufb_0)\Big)\xi+O(\xi^2).
\end{aligned}
\end{equation}
On recalling the  identity
\begin{equation}\label{id}
\ab\times(\bb\times\cb)=(\ab\cdot\bb)\cb-(\ab\cdot\cb)\bb,
\end{equation}
we obtain:
\begin{equation}
\begin{aligned}
(\qb_1\times\qb_2)\times(\qb_1\times\qb_4)=\frac{\sqrt{3}}{2}\ell^2\Big(2\mathfrak{w}_1-\mathfrak{w}_4+\mathfrak{w}_2-2\mathfrak{w}_0 \Big)\ell\pb_1\xi+O(\xi^2),
\end{aligned}
\end{equation}
where we have used the fact that $\pb_4-\pb_2(=-\pb_3-\pb_2)=\pb_1$ and set $\mathfrak{w}=\ufb\cdot\eb_3$.
Now,
\begin{equation}
|\qb_1\times\qb_2|=|\qb_1\times\qb_2|=\frac{\sqrt{3}}{2}\ell^2+O(\xi^2),
\end{equation}
whence
\begin{equation}
|\sin\Big( \Thc_{\pb_1^+}(\xi)\Big)|=|\delta\Thc_{\pb_1^+}|\xi+O(\xi^2)=\frac{2\sqrt{3}}{3\ell}\Big|2\mathfrak{w}_1-\mathfrak{w}_4+\mathfrak{w}_2-2\mathfrak{w}_0  \Big|\xi+O(\xi^2).
\end{equation}
Thus, on recalling that $w=\mathfrak{w}\xi$, we conclude that
\begin{equation}
\delta\Thc_{\pb_1^+}=\frac{2\sqrt{3}}{3\ell}\Big(2w_1-w_4+w_2-2w_0  \Big).
\end{equation}
For a generic C-dihedral angle centered in $\pb_i$, we get
\begin{equation}
\begin{aligned}
&\delta\Thc_{\pb_i^+}(\xb^\ell)=\frac{2\sqrt{3}}{3\ell}[2w(\xb^\ell)-w(\xb^\ell+\ell \pb_{i+1})+w(\xb^\ell+\ell\pb_i-\ell \pb_{i+2})-2w(\xb^\ell+\ell \pb_{i})],\\
&\delta\Thc_{\pb_i^-}(\xb^\ell)=-\frac{2\sqrt{3}}{3\ell}[2w(\xb^\ell)-w(\xb^\ell+\ell \pb_{i+2})+w(\xb^\ell+\ell\pb_i-\ell \pb_{i+1})-2w(\xb^\ell+\ell\pb_i)].
\end{aligned}
\end{equation}		
For the Z-dihedral angle $\Thz_{\pb_1\pb_2}$, we introduce the vector 
$$
\qb_5=\ell\pb_5+(\ufb_5-\ufb_1)\xi,
$$
image of $\pb_5$ under the deformation. We have that 
\begin{equation}
|\sin\Thz_{\pb_1\pb_2}|=\frac{|(\qb_1\times\qb_2)\times(\qb_5\times\qb_1)|}{|\qb_1\times\qb_2||\qb_5\times\qb_1|},
\end{equation}
and
\begin{equation}
\qb_5\times\qb_1=\frac{\sqrt{3}}{2}\ell^2\eb_3+\Big( \ell\pb_5\times(\ufb_1-\ufb_0)-\ell\pb_1\times(\ufb_5-\ufb_1) \Big)\xi+O(\xi^2),
\end{equation}
whence
\begin{equation}
\begin{aligned}
(\qb_1\times\qb_2)\times(\qb_5\times\qb_1)=\frac{\sqrt{3}}{2}\ell^2\eb_3&\times\Big( \ell\pb_5\times(\ufb_1-\ufb_0)-\ell\pb_1\times(\ufb_5-\ufb_1)\\
&+\ell\pb_2\times(\ufb_1-\ufb_0)-\ell\pb_1\times(\ufb_2-\ufb_0) \Big)\xi+O(\xi^2).
\end{aligned}
\end{equation}
Again, on making use of \eqref{id} and recalling that $\pb_5=-\pb_2$, we obtain:
\begin{equation}
(\qb_1\times\qb_2)\times(\qb_5\times\qb_1)=\frac{\sqrt{3}}{2}\ell^2(\mathfrak{w}_5-\mathfrak{w}_1+\mathfrak{w}_2-\mathfrak{w}_0)\xi+O(\xi^2).
\end{equation}
On noticing that $|\qb_5\times\qb_1|=\frac{\sqrt{3}}{2}\ell^2+O(\xi^2)$, we get
\begin{equation}
|\sin\Thz_{\pb_1\pb_2}|=|\delta\Thz_{\pb_1\pb_2}|\xi+O(\xi^2)=\frac{2\sqrt{3}}{3\ell}\Big|\mathfrak{w}_5-\mathfrak{w}_1+\mathfrak{w}_2-\mathfrak{w}_0  \Big|\xi+O(\xi^2),
\end{equation}
whence
\begin{equation}
\delta\Thz_{\pb_1\pb_2}=\frac{2\sqrt{3}}{3\ell}\Big(w_5-w_1+w_2-w_0  \Big).
\end{equation}
For a generic Z-dihedral angle centered in $\pb_i$, we get
\begin{equation}
\begin{aligned}
& \delta\Thz_{\pb_i\pb_{i+1}}(\xb^\ell)=\frac{2\sqrt{3}}{3\ell}[w(\xb^\ell+\ell\pb_i-\ell \pb_{i+1})-w(\xb^\ell+\ell \pb_{i})+w(\xb^\ell+\ell \pb_{i+1})-w(\xb^\ell)],\\
& \delta\Thz_{\pb_i\pb_{i+2}}(\xb^\ell)=\frac{2\sqrt{3}}{3\ell}[w(\xb^\ell+\ell\pb_i-\ell \pb_{i+2})-w(\xb^\ell+\ell \pb_{i})+w(\xb^\ell+\ell \pb_{i+2})-w(\xb^\ell)].
\end{aligned}
\end{equation}

\subsection{Deduction of the continuum limits $\mathcal{U}^{\mathcal{C}}_0$ and $\mathcal{U}^{(s)}_0$}\label{Acont}

In this Appendix we give a justification of \eqref{limenC} and \eqref{limentheta} following the lines outlined in Section \ref{SEC:continuum} for the derivation of the continuum limit of the Z-dihedral energy.

As in Section \ref{SEC:continuum},  in place of a function $w:(L_1(\ell)\cup L_2(\ell))\cap\Omega\to \mathbb{R}$, we consider a twice continuously differentiable function $w:\Omega \to \mathbb{R}$.
We first consider the contribution due to the C-dihedra.
Momentarily, to keep the notation compact, we set
\begin{equation}\label{notgH}
\gb:=\nabla w(\xb^\ell), \qquad \Hb:=\nabla^2 w(\xb^\ell).
\end{equation}
By Taylor's expansion we have
\begin{align*}
-w(\xb^\ell+\ell \pb_{i+1})&=-w(\xb^\ell)-\ell \gb\cdot \pb_{i+1}-\frac 12 \ell^2 \Hb \pb_{i+1}\cdot \pb_{i+1}+o(\ell^2)\\
w(\xb^\ell+\ell \pb_{i}-\ell \pb_{i+2})&=w(\xb^\ell)+\ell \gb\cdot (\pb_i-\pb_{i+2})+\frac 12 \ell^2 \Hb (\pb_i-\pb_{i+2})\cdot (\pb_i-\pb_{i+2})+o(\ell^2)\\
-2w(\xb^\ell+\ell \pb_{i})&=-2w(\xb^\ell)-2\ell \gb\cdot \pb_{i}- \ell^2 \Hb \pb_{i}\cdot \pb_{i}+o(\ell^2).
\end{align*}
Thence, since $\pb_1+\pb_2+\pb_3=\mathbf{0}$, we find from \eqref{C-wedge} that
\begin{align}
\delta\Thc_{\pb_i^+}(\xb^\ell)&=\frac{2\sqrt{3}}{3\ell}[2w(\xb^\ell)-w(\xb^\ell+\ell \pb_{i+1})+w(\xb^\ell+\ell\pb_i-\ell \pb_{i+2})-2w(\xb^\ell+\ell \pb_{i})]\nonumber\\
&=\frac{\sqrt{3}\ell}{3}[-\Hb \pb_{i+1}\cdot \pb_{i+1}+\Hb (\pb_i-\pb_{i+2})\cdot (\pb_i-\pb_{i+2})-2\Hb \pb_{i}\cdot \pb_{i})]+o(\ell)\nonumber
\end{align}
and, by substituting $\pb_{i+2}=-(\pb_{i}+\pb_{i+1})$, we eventually get
\begin{align}
\delta\Thc_{\pb_i^+}(\xb^\ell)&=\frac{\sqrt{3}\ell}{3}[-\Hb \pb_{i+1}\cdot \pb_{i+1}+\Hb (2\pb_i+\pb_{i+1})\cdot (2\pb_i+\pb_{i+1})-2\Hb \pb_{i}\cdot \pb_{i}]+o(\ell)\nonumber\\
&=\frac{4\sqrt{3}\ell}{3}[\Hb \pb_{i}\cdot (\frac 12 \pb_{i}+ \pb_{i+1})]+o(\ell)\nonumber\\
&={2\ell}\Hb \pb_{i}\cdot \pb_{i}^\perp+o(\ell)={2\ell}\partial^2_{\pb_i\pb_i^\perp}w(\xb^\ell)+o(\ell).\label{thC+}
\end{align}
where $\pb_i^\perp$ is defined in \eqref{piperp}. Similarly, we also find that
\begin{equation}\label{thC-}
\delta\Thc_{\pb_i^-}(\xb^\ell)={2\ell}\partial^2_{\pb_i\pb_i^\perp}w(\xb^\ell)+o(\ell).
\end{equation}
With the same steps taken in the study of the Z-dihedral energy we now derive the limit of the C-dihedral energy.
With the expressions of the change of the C-dihedra \eqref{thC+} and \eqref{thC-}, we can rewrite the C-dihedral energy,
see \eqref{enC}, as
\begin{align*}
\mathcal{U}^{\mathcal{C}}_\ell&=\frac{1}{2} k^{\mathcal{C}} \, \sum_{\xb^\ell\in L_2(\ell)\cap\Omega}
\sum_{i=1}^3 \Bigg(\delta\Thc_{\pb_i^+}(\xb^\ell)\Bigg)^2+ \Bigg(\delta\Thc_{\pb_i^-}(\xb^\ell)\Bigg)^2
\nonumber\\
&=\frac{1}{2} 8\ell^2 k^{\mathcal{C}} \, \sum_{\xb^\ell\in L_2(\ell)\cap\Omega}
\sum_{i=1}^3 \Big(\partial^2_{\pb_i\pb_i^\perp}w(\xb^\ell)\Big)^2+o(\ell^2)
\nonumber\\
&=\frac{1}{2} \frac{16\sqrt{3}}{9} k^{\mathcal{C}} \, \sum_{\xb^\ell\in L_2(\ell)\cap\Omega}
\sum_{i=1}^3 \Big(\partial^2_{\pb_i\pb_i^\perp}w(\xb^\ell)\Big)^2|E^\ell(\xb^\ell)|+o(\ell^2)
\nonumber\\
&=\frac{1}{2} \frac{16\sqrt{3}}{9} k^{\mathcal{C}} \int_\Omega W^\Cc_\ell(\xb)\,d\xb+o(\ell^2),
\end{align*}
where the function
$$
W^\Cc_\ell(\xb):=\sum_{\xb^\ell\in L_2(\ell)\cap\Omega}
\sum_{i=1}^3 \Big(\partial^2_{\pb_i\pb_i^\perp}w(\xb^\ell)\Big)^2 \chi_{E^\ell(\xb^\ell)}(\xb)
$$
converges to $\sum_{i=1}^3 \Big(\partial^2_{\pb_i\pb_i^\perp}w\Big)^2$, as $\ell$ goes to zero.
Thus, 
\begin{equation}\label{limenC2}
\lim_{\ell\to 0}\mathcal{U}^{\mathcal{C}}_\ell=\frac{1}{2} \frac{16\sqrt{3}}{9} k^{\mathcal{C}}\int_\Omega \sum_{i=1}^3 \Big(\partial^2_{\pb_i\pb_i^\perp}w\Big)^2\,d\xb
=:\mathcal{U}^{\mathcal{C}}_0(w),
\end{equation}
which is \eqref{limenC}.

\begin{figure}[h]
	\centering
	\includegraphics[scale=1.2]{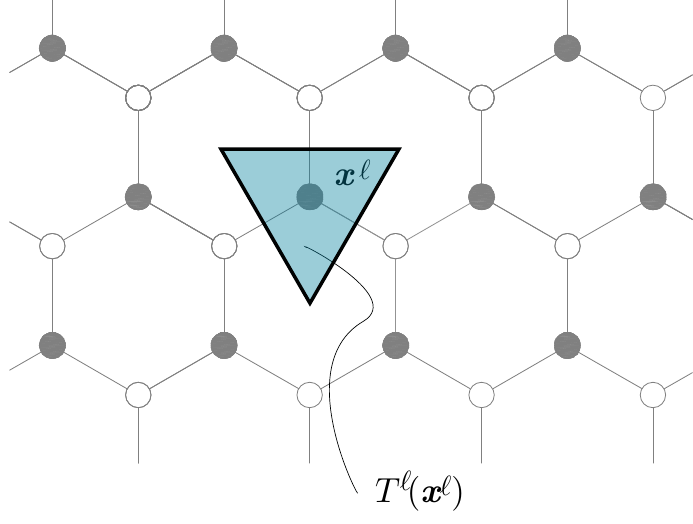}
	\caption{Triangulation  $T^\ell(\xb^\ell)$}\label{triangle}
\end{figure}

We now compute the limit of the self-energy.
By Taylor's expansion, with the notation introduced in \eqref{notgH}, and taking into account that
$\pb_1+\pb_2+\pb_3=\mathbf{0}$, we find
\begin{align}
\sum_i\delta\vartheta_i(\xb^\ell)&=-\frac{3\sqrt{3}}{\ell^2}[\frac 13 \sum_{i=1}^3w(\xb^\ell+\ell\pb_i)-w(\xb^\ell)]^2\nonumber\\
&=-\frac{3\sqrt{3}}{\ell^2}\frac{\ell^4}{36} ( \Hb\pb_1\cdot \pb_1+\Hb\pb_2\cdot \pb_2+\Hb\pb_3\cdot \pb_3)^2+o(\ell^2)\nonumber\\
&=-\frac{3\sqrt{3}}{\ell^2}\frac{\ell^4}9( \Hb\pb_1\cdot \pb_1+\Hb\pb_2\cdot \pb_2+\Hb\pb_1\cdot \pb_2)^2+o(\ell^2)\nonumber\\
&=-\frac{\sqrt{3}}{3} {\ell^2}\Big( \partial^2_{\pb_1 \pb_1}w(\xb^\ell)+\partial^2_{\pb_2 \pb_2}w(\xb^\ell)+\partial^2_{\pb_1 \pb_2}w(\xb^\ell)\Big)^2+o(\ell^2).\label{sumthi}
\end{align}
With this expression the self-energy \eqref{selfen1} takes the form
\begin{align}
\mathcal{U}^{(s)}_\ell
&= \sum_{\xb^\ell\in (L_1(\ell)\cup L_2(\ell))\cap\Omega}   \tau_0 \sum_{i = 
	1}^3\delta \vartheta_i (\xb^\ell) \nonumber \\		
&= -\frac{\sqrt{3}}3\tau_0 {\ell^2}\sum_{\xb^\ell\in (L_1(\ell)\cup L_2(\ell))\cap\Omega}   
\Big( \partial^2_{\pb_1 \pb_1}w(\xb^\ell)+\partial^2_{\pb_2 \pb_2}w(\xb^\ell)+\partial^2_{\pb_1 \pb_2}w(\xb^\ell)\Big)^2+o(\ell^2). \nonumber 
\end{align}
Since the sum is over the points of both lattices, whereas in the previous cases the sum was only over the nodes of $L_2(\ell)$, we cannot use the hexagons $E^{\ell}(\xb^\ell)$ earlier introduced. Let $T^\ell(\xb^\ell)$ be the triangle 
centered at $\xb^\ell$ of side $\sqrt{3}\ell$ as depicted in Figure \ref{triangle}.

Let
$$
W^\vartheta_\ell(\xb):=\sum_{\xb^\ell\in (L_1(\ell)\cup L_2(\ell))\cap\Omega}
\Big(\partial^2_{\pb_1 \pb_1}w(\xb^\ell)+\partial^2_{\pb_2 \pb_2}w(\xb^\ell)+\partial^2_{\pb_1 \pb_2}w(\xb^\ell)
\Big)^2 \chi_{T^\ell(\xb^\ell)}(\xb)
$$
and note that the area of $T^\ell(\xb^\ell)$ is $|T^\ell(\xb^\ell)|=\frac{3\sqrt{3}}4\ell^2$.
The self-energy rewrites as
\begin{align}
\mathcal{U}^{(s)}_\ell
&= -\frac{4}9\tau_0 \sum_{\xb^\ell\in (L_1(\ell)\cup L_2(\ell))\cap\Omega}   
\Big( \partial^2_{\pb_1 \pb_1}w(\xb^\ell)+\partial^2_{\pb_2 \pb_2}w(\xb^\ell)+\partial^2_{\pb_1 \pb_2}w(\xb^\ell)\Big)^2|T^\ell(\xb^\ell)| \nonumber \\
&\quad+o(\ell^2) \nonumber \\
&= -\frac{4}9\tau_0 \int_\Omega W^\vartheta_\ell(\xb)\, d\xb+o(\ell^2).\nonumber
\end{align}
Since $W^\vartheta_\ell$ converges to $( \partial^2_{\pb_1 \pb_1}w+\partial^2_{\pb_2 \pb_2}w+\partial^2_{\pb_1 \pb_2}w\Big)^2$ as $\ell$ goes to zero, we find
\begin{align}
\lim_{\ell\to 0}\mathcal{U}^{(s)}_\ell=-\frac{4}9\tau_0 \int_\Omega ( \partial^2_{\pb_1 \pb_1}w+\partial^2_{\pb_2 \pb_2}w+\partial^2_{\pb_1 \pb_2}w\Big)^2\, d\xb
=:\mathcal{U}^{(s)}_0(w),\label{limentheta2}
\end{align}
which is \eqref{limentheta}.

\end{document}